\documentclass[10pt,a4paper,reqno]{amsart}
\usepackage{}
\usepackage{acronym}
\usepackage{amsfonts}
\usepackage{amsmath}
\usepackage{amssymb}
\usepackage{amsthm}
\usepackage{bbm}
\usepackage{bm}
\usepackage{braket}
\usepackage{cite}
\usepackage{color}
\usepackage{geometry}
\usepackage{graphicx}
\usepackage{hyperref}
\usepackage{mathdots}
\usepackage{mathrsfs}
\usepackage{mathtools}
\usepackage{setspace}
\usepackage{stmaryrd}
\usepackage{subcaption}
\usepackage{tikz}
\usepackage{txfonts}

\newgeometry{top=1.5cm,bottom=1.5cm,outer=1.5cm,inner=1.5cm}

\newcommand{\bse}{\begin{subequations}}
\newcommand{\ese}{\end{subequations}}

\newtheorem{theorem}{Theorem}[section]

\newtheorem{example}[theorem]{Example}
\numberwithin{equation}{section}

\DeclareMathOperator{\tr}{tr}
\DeclareMathOperator{\diag}{diag}

\def\undertilde#1{\mathord{\vtop{\ialign{##\crcr
$\hfil\displaystyle{#1}\hfil$\crcr\noalign{\kern1.5pt\nointerlineskip}
$\hfil\widetilde{}\hfil$\crcr\noalign{\kern-6.5pt}}}}}
\def\underhat#1{\mathord{\vtop{\ialign{##\crcr
$\hfil\displaystyle{#1}\hfil$\crcr\noalign{\kern1.5pt\nointerlineskip}
$\hfil\widehat{}\hfil$\crcr\noalign{\kern-6.5pt}}}}}
\def\underbar#1{\mathord{\vtop{\ialign{##\crcr
$\hfil\displaystyle{#1}\hfil$\crcr\noalign{\kern1.5pt\nointerlineskip}
$\hfil\bar{}\hfil$\crcr\noalign{\kern-6.5pt}}}}}

\newcommand{\rD}{\mathrm{D}}
\newcommand{\rd}{\mathrm{d}}
\newcommand{\re}{\mathrm{e}}
\newcommand{\ri}{\mathrm{i}}

\newcommand{\cN}{\mathcal{N}}

\newcommand{\bLd}{\mathbf{\Lambda}}
\newcommand{\tbLd}{{}^{t\!}\boldsymbol{\Lambda}}

\newcommand{\bO}{\boldsymbol{O}}
\newcommand{\tbO}{{}^{t\!}\boldsymbol{O}}
\newcommand{\bOa}{\boldsymbol{\Omega}}
\newcommand{\tbOa}{{}^{t\!}\boldsymbol{\Omega}}
\newcommand{\bC}{\boldsymbol{C}}
\newcommand{\tbC}{{}^{t\!}\boldsymbol{C}}
\newcommand{\bU}{\boldsymbol{U}}
\newcommand{\tbU}{{}^{t\!}\boldsymbol{U}}
\newcommand{\bu}{\boldsymbol{u}}
\newcommand{\tbu}{{}^{t\!}\boldsymbol{u}}

\newcommand{\bv}{\boldsymbol{v}}
\newcommand{\tbv}{{}^{t\!}\boldsymbol{v}}
\newcommand{\bc}{\boldsymbol{c}}
\newcommand{\tbc}{{}^{t\!}\boldsymbol{c}}
\newcommand{\bA}{\boldsymbol{A}}

\newcommand{\bD}{\boldsymbol{D}}

\newcommand{\bG}{\boldsymbol{G}}

\newcommand{\bR}{\boldsymbol{R}}
\newcommand{\bS}{\boldsymbol{S}}

\newcommand{\bM}{\boldsymbol{M}}
\newcommand{\bI}{\boldsymbol{I}}
\newcommand{\bV}{\boldsymbol{V}}

\newcommand{\ld}{\lambda}

\newcommand{\oa}{\omega}
\newcommand{\Oa}{\Omega}

\newcommand{\bP}{\boldsymbol{P}}
\newcommand{\bQ}{\boldsymbol{Q}}

\newcommand{\br}{\boldsymbol{r}}
\newcommand{\tbr}{{}^{t\!}\boldsymbol{r}}
\newcommand{\bs}{\boldsymbol{s}}
\newcommand{\tbs}{{}^{t\!}\boldsymbol{s}}
\newcommand{\bK}{\boldsymbol{K}}

\newcommand{\tbuscr}{{}^{t\!}\mathscr{U}}

\newcommand{\tbcscr}{{}^{t\!}\mathscr{C}}
\newcommand{\bcscr}{\mathscr{C}}

\acrodef{1D}[1D]{one-dimensional}
\acrodef{2D}[2D]{two-dimensional}
\acrodef{2DTL}[2DTL]{two-dimensional Toda lattice}
\acrodef{3D}[3D]{three-dimensional}
\acrodef{ABS}[ABS]{Adler--Bobenko--Suris}
\acrodef{bDT}[bDT]{binary Darboux transform}
\acrodef{BT}[BT]{B\"acklund transform}
\acrodef{BSQ}[BSQ]{Boussinesq}
\acrodef{CAC}[CAC]{consistency-around-the-cube}
\acrodef{DT}[DT]{Darboux transform}
\acrodef{DL}[DL]{direct linearisation}
\acrodef{DLT}[DLT]{direct linearising transform}
\acrodef{DDeltaE}[D$\Delta$E]{differential-difference equation}
\acrodef{FX}[FX]{Fordy--Xenitidis}
\acrodef{GD}[GD]{Gel'fand--Dikii}
\acrodef{KP}[KP]{Kadomtsev--Petviashvili}
\acrodef{KdV}[KdV]{Korteweg--de Vries}
\acrodef{HM}[HM]{Hirota--Miwa}
\acrodef{MDC}[MDC]{multi-dimensional consistency}
\acrodef{NQC}[NQC]{Nijhoff--Quispel--Capel}
\acrodef{ODE}[ODE]{ordinary differential equation}
\acrodef{ODeltaE}[O$\Delta$E]{ordinary difference equation}
\acrodef{PDE}[PDE]{partial differential equation}
\acrodef{PDeltaE}[P$\Delta$E]{partial difference equation}
\acrodef{RHP}[RHP]{Riemann--Hilebert problem}
\acrodef{sG}[sG]{sine--Gordon}
\acrodef{YB}[YB]{Yang--Baxter}

\title[Linear integral equations and two-dimensional Toda systems]{Linear integral equations and two-dimensional Toda systems}
\author{Yue Yin}
\address[YY]{School of Mathematical Sciences \\
East China Normal University \\ 500 Dongchuan Road \\ Shanghai 200241 \\ People's Republic of China}
\author{Wei Fu}
\address[WF]{School of Mathematical Sciences and Shanghai Key Laboratory of Pure Mathematics and Mathematical Practice \\
East China Normal University \\ 500 Dongchuan Road \\ Shanghai 200241 \\ People's Republic of China}

\begin{document}

\begin{abstract}
The direct linearisation framework is presented for the two-dimensional Toda equations associated with the infinite-dimensional Lie algebras $A_\infty$, $B_\infty$ and $C_\infty$,
as well as the Kac--Moody algebras $A_{r}^{(1)}$, $A_{2r}^{(2)}$, $C_{r}^{(1)}$ and $D_{r+1}^{(2)}$ for arbitrary integers $r\in\mathbb{Z}^+$,
from the aspect of a set of linear integral equations in a certain form.
Such a scheme not only provides a unified perspective to understand the underlying integrability structure,
but also induces the direct linearising type solution potentially leading to the universal solution space, for each class of the two-dimensional Toda system.
As particular applications of this framework to the two-dimensional Toda lattices, we rediscover the Lax pairs and the adjoint Lax pairs
and simultaneously construct the generalised Cauchy matrix solutions.
\end{abstract}


\maketitle

\section{Introduction}\label{S:Intro}

In the modern theory of integrable systems, the notion of integrability of nonlinear equations often refers to the property that
a differential/difference equation is exactly solvable under an initial-boundary condition, which in many cases allows us to construct explicit solutions.
Motivated by this, many mathematical methods were invented and developed to search for explicit solutions of nonlinear models,
for instance, the inverse scattering transform, the Darboux transform, Hirota's bilinear method, as well as the algebro-geometric method, etc.,
see e.g. the monographs \cite{AC91,Mat91,Hir04,NMPZ84}.
These techniques not only explained the nonlinear phenomena such as solitary waves and periodic waves in nature mathematically,
but also motivated the discovery of a huge number of nonlinear equations that possess ``nice'' algebraic and geometric properties.

Among integrable systems, a very typical model is the (2+1)-dimensional equation
\begin{align}\label{A:NL}
\partial_1\partial_{-1}\varphi_n=\re^{\varphi_n-\varphi_{n-1}}-\re^{\varphi_{n+1}-\varphi_n},
\end{align}
in which $\varphi_n=\varphi(x_1,x_{-1},n)$ is the potential function of two continuous time variables $x_1$ and $x_{-1}$ and one discrete spatial variable $n$.
Here $\partial_1$ and $\partial_{-1}$ stand for the partial derivatives with respect to $x_1$ and $x_{-1}$, respectively.
This equation was proposed by Mikhailov \cite{Mik79} in 1979, as an integrable extension of the famous \ac{1D} Toda lattice (cf. \cite{Tod67})
\begin{align*}
\partial_1^2\varphi_n=\re^{\varphi_n-\varphi_{n-1}}-\re^{\varphi_{n+1}-\varphi_n},
\end{align*}
and thus it is often referred to as the \ac{2D} Toda equation.
Equation \eqref{A:NL} can alternatively be written as a (1+1)-dimensional system composed of an infinite number of components as follows (see e.g. \cite{JM83}):
\begin{align}\label{A:Cartan}
\partial_1\partial_{-1}\left(
\begin{array}{c}
\vdots\\
\theta_{-1}\\
\boxed{\theta_0} \\
\theta_1 \\
\vdots
\end{array}
\right)=-
\begin{pmatrix}
\ddots&\ddots& \ddots \\
&-1 & 2 & -1 \\
& & -1 & \boxed{2} & -1 \\
& & & -1 & 2 & -1 \\
& & & & \ddots & \ddots & \ddots
\end{pmatrix}
\left(
\begin{array}{c}
\vdots \\
\re^{-\theta_{-1}} \\
\boxed{\re^{-\theta_{0}}} \\
\re^{-\theta_1} \\
\vdots
\end{array}
\right),
\end{align}
where $\theta_n\doteq\varphi_{n-1}-\varphi_n$, and the boxes denote the respective locations of the central elements of an infinite vector and an infinite matrix,
namely the $0$th component and the $(0,0)$-entry.

The \ac{2D} Toda equation \eqref{A:NL} is closely related to the representation theory.
In a series of papers from various authors, see e.g. \cite{MOP83,Wil81,JM83,UT84,NW97},
a huge class of \ac{2D} Toda-type equations were obtained based on the classification of Lie algebras.
All these equations were proven integrable from various perspectives.
The characteristic of Lie algebra also reflects in the nonlinear equations themselves.
For instance, the entries of the Cartan matrix of $A_\infty$ appears as the coefficients in \eqref{A:Cartan},
and hence, equation \eqref{A:NL} is often referred to as the \ac{2DTL} of $A_\infty$-type.
The \ac{2D} Toda-type equations in other classes also take similar multi-component form with regard to their respective Cartan matrices.
In the whole classification of the \ac{2DTL}s, only the equations associated with the Kac--Moody algebras $A_1^{(1)}$ and $A_2^{(2)}$,
namely the sinh--Gordon equation
\begin{align}\label{sG}
\partial_1\partial_{-1}\varphi_0=\re^{2\varphi_0}-\re^{-2\varphi_0}
\end{align}
and the Tzitzeica equation
\bse\label{Tz}
\begin{align}\label{Tz1}
\partial_1\partial_{-1}\varphi_0=\re^{\varphi_0}-\re^{-2\varphi_0},
\end{align}
or alternatively,
\begin{align}\label{Tz2}
\partial_1\partial_{-1}\varphi_0=\re^{2\varphi_0}-\re^{-\varphi_0},
\end{align}
\ese
can be written in scalar form as \ac{2D} models, in addition to the \ac{3D} equation \eqref{A:NL}.

We are intended to study the \ac{2D} Toda-type equations within the so-called \ac{DL} scheme.
The \ac{DL} was first proposed by Fokas, Ablowitz and Santini \cite{FA81,FA83,SAF84} to solve initial value problems of nonlinear partial differential equations
including the \ac{KdV} and \ac{KP} equations, as a kind of generalisation of the dressing method or the $\bar{\partial}$ method.
In addition to solving a nonlinear equation, the \ac{DL} approach also simultaneously provides an insight into understanding the underlying structures of integrable equations.
This is reflected in many aspects including searching for integrable discretisation of differential equations (see e.g. the review papers \cite{QNCL84,NCW85}),
constructing integrability characteristics of nonlinear equations such as Lax pairs \cite{NPCQ92,ZZN12,Fu20},
commuting symmetries \cite{Fu18b}, and master symmetries \cite{NRGO01,NHJ00}, etc.
Recently, one of the authors and his collaborator further developed the \ac{DL} and established the connection between linear integral equations and the Lie algebras \cite{Fu17a,Fu18b,Fu21a}, which extended the range of application of the \ac{DL}.
The main idea behind the \ac{DL} approach, according to Fokas and Ablowitz \cite{FA81},
is to construct solutions of a nonlinear equation, by solving its relevant linear integral equation.
This brings the benefit that no additional conditions are imposed at the beginning, and thus,
a large class of solutions are obtainable as various degenerations from a so-called direct linearising solution,
which takes the form of an integral defined on an arbitrary integration domain through an arbitrary integration measure in terms of the spectral variable(s).
In this sense, the establishment of the \ac{DL} scheme of the \ac{2DTL}s potentially provides a path towards their respective `universal' solution spaces.

In the present paper, we consider the \ac{2D} Toda-type equations associated with the infinite-dimensional algebras $A_{\infty}$, $B_\infty$, $C_{\infty}$,
as well as the Kac--Moody Lie algebras $A_{r}^{(1)}$, $A_{2r}^{(2)}$, $C_r^{(1)}$ and $D_{r+1}^{(2)}$, for arbitrary integers $r\in\mathbb{Z}^+$.
Our starting point is a set of linear integral equations (namely a linear integral equation and its adjoint) taking the form of \eqref{Linear}.
By selecting suitable plane wave factors and kernel, i.e. \eqref{PWF} and \eqref{Kernel},
the $A_\infty$-type {2D} Toda system is recovered as the nonlinearisation of the wave function of either of the linear integral equations, in the language of infinite matrix.
From such a set of linear integral equations, we are able to construct its direct linearising type solution,
and also the integrability characteristics including the Lax pair, the adjoint Lax pair as well as the tau function, for the \ac{2DTL} of $A_\infty$-type.
By performing various constraints on the integration measure in terms of the spectral parameters,
we obtain the respective \ac{DL} schemes for the \ac{2DTL}s of  $B_\infty$-, $C_{\infty}$-, $A_{r}^{(1)}$-, $A_{2r}^{(2)}$-, $C_r^{(1)}$- and $D_{r+1}^{(2)}$-types
and subsequently the corresponding direct linearising solutions, the linear problems, and also the tau functions.
These form a unified framework for the integrability of the \ac{2DTL}s,
which generalises the \ac{DL} approach proposed by Fokas and Ablowitz to the \ac{2D} Toda systems of various types,
and simultaneously provides an insight into the results obtained in the literature (cf. e.g. \cite{FG83,UT84,NW97}) from a different perspective.

In addition, we also discuss the so-called generalised Cauchy matrix solutions for the various \ac{2D} Toda-type equations.
This is realised by taking a particular integration measure possessing an arbitrary fixed number of higher-order singularities, with respect to the spectral variable(s).
As a consequence, the linear integral equation reduces to a linear algebraic equation,
which allows us to establish the general formula of the Cauchy matrix type solution, for each \ac{2DTL}.
This generalises the results obtained in \cite{NAH09,ZZ13,XZZ14} from $A_1^{(1)}$-type to other types.

The paper is organised as follows. The formal structure of our approach is introduced in section \ref{S:DL}, illustrated in the language of infinite matrix.
In section \ref{S:A}, we establish \ac{DL} scheme of the \ac{2DTL} of $A_{\infty}$-type,
from which we construct its direct linearising solution, tau function, and associated linear problems.
The \ac{DL} schemes and the relevant results of the \ac{2DTL}s of $B_\infty$-, $C_{\infty}$-,
as well as $A_{r}^{(1)}$-, $A_{2r}^{(2)}$-, $C_r^{(1)}$- and $D_{r+1}^{(2)}$-types are discussed in sections \ref{S:BC} and \ref{S:Reduc}, respectively.
In section \ref{S:Sol}, general formulae of the Cauchy matrix type solutions for the \ac{2DTL}s are presented.

\section{Preliminaries}\label{S:DL}

\subsection{Infinite matrices and vectors}

Suppose that an infinite matrix of size $\infty\times\infty$ is given by
\begin{align*}
\bU=
\begin{pmatrix}
& \vdots & \vdots & \vdots \\
\cdots & U_{-1,-1} & U_{-1,0} & U_{-1,1} & \cdots\\
\cdots & U_{0,-1} & \boxed{U_{0,0}} & U_{0,1} & \cdots\\
\cdots & U_{1,-1} & U_{1,0} & U_{1,1} & \cdots\\
& \vdots & \vdots & \vdots
\end{pmatrix}.
\end{align*}
We define the operations for these objects by following the same rules in the finite-dimensional case,
namely for arbitrary infinite matrices $\bU=(U_{i,j})_{\infty\times\infty}$ and $\bV=(V_{i,j})_{\infty\times\infty}$,
the addition $\bU+\bV$, the scalar multiplication $p\,\bU$ and the matrix multiplication $\bU\bV$ give rise to infinite matrices of size $\infty\times\infty$,
and their respective $(i,j)$-entries of the addition $\bU+\bV$, the scalar multiplication $p\,\bU$ and the matrix multiplication $\bU\bV$ are defined as
$U_{i,j}+V_{i,j}$, $p\,U_{i,j}$ and $\sum_{i'\in\mathbb{Z}}U_{i,i'}V_{i',j}$\footnote{
The multiplication is a formal definition.
In general, the infinite summation could raise the issue of divergence;
however, we only deal with convergent infinite summation throughout the paper.
}.
Similarly, an infinite column vector and its transpose are given by
\begin{align*}
\bu={}^{t\!}(\cdots,u_{-1},\boxed{u_{0}},u_{1},\cdots) \quad \hbox{and} \quad \tbu=(\cdots,u_{-1},\boxed{u_{0}},u_{1},\cdots),
\end{align*}
respectively. Here we note that we adopt the notation ${}^{t\!}(\,\cdot\,)$ for the transpose of a matrix or a vector.
For arbitrary infinite column vectors $\bu=(u_i)_{\infty\times 1}$ and $\bv=(v_i)_{\infty\times 1}$,
$\bu+\bv$ and $p\,\bu$ are infinite column vectors whose $i$th components are defined as $u_i+v_i$ and $p\,u_i$, respectively,
$\bu\tbv$ is an $\infty\times\infty$ infinite matrix having its $(i,j)$-entry $u_iv_j$, and $\tbu\bv$ is a scalar defined by $\sum_{i\in\mathbb{Z}}u_iv_i$.
The multiplication between an infinite matrix and an infinite vector is defined in the same way, i.e.
$\bU\bv$ is an infinite column vector with its $i$th component given by $\sum_{j\in\mathbb{Z}}U_{i,j}v_j$,
and $\tbv\bU$ is an infinite row vector with its $j$th component defined as $\sum_{i\in\mathbb{Z}}v_iU_{i,j}$.

Next, we introduce some special infinite matrices and vectors. The projection matrix is defined as
\begin{align*}
\bO\doteq
\begin{pmatrix}
& \vdots & \vdots & \vdots \\
\cdots & 0 & 0 & 0 & \cdots \\
\cdots & 0 & \boxed{1} & 0 & \cdots \\
\cdots & 0 & 0 & 0 & \cdots \\
& \vdots & \vdots & \vdots
\end{pmatrix}.
\end{align*}
It is easily verified that the projection matrix $\bO$ satisfies the properties
\begin{align*}
(\bO\bU)^{(i,j)}=\delta_{i,0}U_{0,j} \quad \hbox{and} \quad (\bU\bO)^{(i,j)}=U_{i,0}\delta_{0,j}, \quad \hbox{where} \quad
\delta_{i,j}=
\left\{
\begin{array}{ll}
1, & i=j, \\
0, & i\neq j,
\end{array}
\right. \quad
\forall i,j\in\mathbb{Z},
\end{align*}
as well as $\bO^2=\bO$ and $\tbO=\bO$.
Here the notation $(\,\cdot\,)^{(i,j)}$ stands for the $(i,j)$-entry of an infinite matrix.
The index-raising matrix $\bLd$ and its transpose $\tbLd$ are defined as
\begin{align*}
\bLd\doteq
\begin{pmatrix}
\ddots & \ddots \\
& 0 & 1 \\
& & \boxed{0} & 1 \\
& & & 0 & \ddots \\
& & & & \ddots
\end{pmatrix}
\quad \hbox{and} \quad
\tbLd\doteq
\begin{pmatrix}
\ddots \\
\ddots & 0 \\
& 1 & \boxed{0} \\
& & 1 & 0 & \\
& & & \ddots & \ddots
\end{pmatrix},
\end{align*}
respectively. Multiplying $\bU$ by $\bLd$ from the left and $\tbLd$ from the right, respectively, one obtains
\begin{align*}
(\bLd\,\bU)^{(i,j)}=U_{i+1,j} \quad \hbox{and} \quad (\bU\,\tbLd)^{(i,j)}=U_{i,j+1},
\end{align*}
namely, the operation of the index raising matrices $\bLd$ and $\tbLd$ raise the row and column indices by $1$, respectively.
In our approach we also need the column and row infinite vectors
\begin{align}\label{bc}
\bc(k)\doteq{}^{t\!}(\cdots,k^{-1},\boxed{1},k,\cdots) \quad \hbox{and} \quad
\tbc(k')\doteq(\cdots,k'^{-1},\boxed{1},k',\cdots),
\end{align}
composed of monomials of $k$ and $k'$.
The operation of $\bLd$ and $\tbLd$ on these two infinite vectors give rise to relations
\begin{align}\label{bcDyn}
\bLd\bc(k)=k\,\bc(k) \quad \hbox{and} \quad \tbc(k')\tbLd=k'\tbc(k'),
\end{align}
or equivalently, $[\bLd\bc(k)]^{(i)}=k^{i+1}$ and $[\tbc(k')\tbLd]^{(i)}=k'^{i+1}$,
where the notation $(\,\cdot\,)^{(i)}$ denotes the $i$th component of an infinite vector.
In addition, it is also proven that
\begin{align*}
\tbc(k')\tbLd^{j}\bO\bLd^i\bc(k)=k^ik'^j.
\end{align*}

The trace of an arbitrary infinite matrix $\bU$ is defined as
\begin{align*}
\tr\bU=\sum_{i\in\mathbb{Z}}U_{i,i},
\end{align*}
and it possesses the property
\begin{align}\label{Commute}
\tr(\bU\bV)=\tr(\bV\bU).
\end{align}
The determinant of an infinite matrix is defined through
\begin{align*}
\ln(\det\bU)=\tr(\ln\bU).
\end{align*}
In practice we only deal with the determinant of a very special infinite matrix in the form of $1+*$ (where $1$ denotes the identity infinite matrix) in our paper,
and thus, we are allowed to reformulate the determinant as
\begin{align*}
\det(1+*)=\exp\{\tr[\ln(1+*)]\}=\exp\left\{\sum_{i=1}^\infty\frac{(-1)^{i-1}}{i}\tr\left(*^i\right)\right\},
\end{align*}
which is meaningful as long as the infinite matrix $*$ guarantees convergent expansion.
The determinant satisfies the property
\begin{align}\label{Rank1}
\det(1+\bU\bO\bV)=1+(\bV\bU)^{(0,0)},
\end{align}
namely the well-known Weinstein--Aronszajn formula for the rank $1$ case.

\subsection{Infinite matrix representation of the linear integral equations}

We start with introducing the notations for the key objects in the linear integral equations.
The infinite column vector $\bu(k)$ and the infinite row vector $\tbv(k')$ are the wave functions,
whose components are functions of the dynamical variables (which could be either continuous or discrete arguments),
relying on the spectral parameters $k$ and $k'$, respectively.
For the plane wave factors, we adopt the notations $\rho(k)$ and $\sigma(k')$, which are scalars dependent on the dynamical variables,
as well as their respective spectral parameters $k$ and $k'$.
The Cauchy kernel $\Oa(k,k')$ is a scalar function of only the spectral parameters $k$ and $k'$.
The linear integral equations we consider in our scheme are as follows:
\bse\label{Linear}
\begin{align}
&\bu(k)+\iint_D\rd\zeta(l,l')\rho(k)\Oa(k,l')\sigma(l')\bu(l)=\rho(k)\bc(k), \label{Lineara} \\
&\tbv(k')+\iint_D\rd\zeta(l,l')\rho(l)\Oa(l,k')\sigma(k')\tbv(l')=\tbc(k')\sigma(k'), \label{Linearb}
\end{align}
\ese
where $k$ and $k'$ (or $l$ and $l'$) are the spectral parameters in the complex field $\mathbb{C}$,
$\rd\zeta$ is the integration measure, and $D$ is the integration domain.
In the paper, we refer to equation \eqref{Linearb} as the `adjoint' of \eqref{Lineara}.
This set of equations, i.e. \eqref{Linear}, is the starting point for the \ac{DL} approach,
and at this stage, there is no restriction (such as restricted domain and measure, etc.) on \eqref{Linear},
except that we require that the homogeneous equation corresponding to any of \eqref{Lineara} and \eqref{Linearb} has only zero solution.

Next, we consider the infinite matrix representation of the linear integral equations.
If we introduce the infinite matrices $\bOa$ and $\bU$, which are defined by
\begin{align}\label{Omega}
\Oa(k,k')\doteq\tbc(k')\bOa\bc(k)
\end{align}
and
\begin{align}\label{Potential}
\bU\doteq\iint_D\rd\zeta(k,k')\bu(k)\tbc(k')\sigma(k'),\quad \hbox{or equivalently} \quad
\bU\doteq\iint_D\rd\zeta(k,k')\rho(k)\bc(k)\tbv(k'),
\end{align}
respectively, and replace the kernel $\Oa(k,k')$ with the help of \eqref{Omega} and \eqref{Potential},
the linear integral equations in \eqref{Linear} are reformulated as follows:
\bse\label{uv}
\begin{align}
&\bu(k)=(1-\bU\bOa)\bc(k)\rho(k), \label{u} \\
&\tbv(k')=\sigma(k')\tbc(k')(1-\bOa\bU). \label{v}
\end{align}
\ese
Notice the first definition of $\bU$ given in \eqref{Potential}. We immediately derive from \eqref{u} the infinite matrix relation\footnote{
Similarly, one is able to derive $\bU=\bC(1-\bOa\bU)$ from \eqref{v} by making use of the second definition in \eqref{Potential}.
This leads to another relation $\bU=(1+\bC\bOa)^{-1}\bC$, which is equivalent to \eqref{U}.
This shows why the two definitions in \eqref{Potential} are equivalent.
}
\begin{align}\label{U}
\bU=(1-\bU\bOa)\bC, \quad \hbox{or alternatively} \quad \bU=\bC(1+\bOa\bC)^{-1},
\end{align}
where the infinite matrix $\bC$ is defined as
\begin{align}\label{C}
\bC\doteq\iint_D\rd\zeta(k,k')\rho(k)\bc(k)\tbc(k')\sigma(k').
\end{align}
In our framework, we also define the formal tau function as
\begin{align}\label{tau}
\tau\doteq\det(1+\bOa\bC),
\end{align}
or equivalently $\tau\doteq\det(1+\bC\bOa)$.

The infinite matrix $\bU$, the infinite vectors $\bu(k)$ and $\tbv(k')$ and the tau function $\tau$ are the key quantities in the our framework.
Once the kernel and the plane wave factors in the linear integral equations are fully determined,
we are able to construct nonlinear equations and its integrability characteristics based on these quantities.
More concretely, the entries of the infinite matrix $\bU$ (or their combinations) form potentials of the resulting nonlinear integrable equations;
the components of the infinite vectors $\bu(k)$ and $\tbv(k')$ play roles of the eigenfunctions of the associated linear problems (i.e. Lax pairs and the adjoint ones);
while the tau function is the potential for the corresponding bilinear equations.
To put it another way, \eqref{Potential} subject to \eqref{Linear} provides us with a general solution to the resulting nonlinear integrable equations within our scheme,
which contains a large class of exact solutions as special cases.

\section{2DTL associated with $A_\infty$}\label{S:A}

\subsection{Closed-form nonlinear equations}

We start with the \ac{2DTL} of $A_\infty$-type in this section from the formal structure of the \ac{DL}.
To establish the scheme, we select the plane wave factors and the Cauchy kernel given by
\begin{align}\label{PWF}
\rho_n(k)=\re^{kx_{1}+k^{-1}x_{-1}}k^n \quad \hbox{and} \quad \sigma_n({k'})=\re^{k^{\prime}x_1+k^{\prime -1}x_{-1}}(-k')^{-n}
\end{align}
as well as
\begin{align}\label{Kernel}
\Oa(k,k')=\frac{1}{k+k'},
\end{align}
respectively, where $x_1$ and $x_{-1}$ are the continuous temporal variables, $n$ is the discrete spatial variable.
Since we need to consider the discrete dynamical evolutions in our scheme,
from now on we adopt the suffix $n$ for various quantities,
in order to explicitly illustrate how these quantities evolves with respect to $n$, when it is necessary.

Notice the definition of $\bC_n$, namely \eqref{C}, for given plane wave factors in \eqref{PWF}.
We can prove that $\bC_n$ satisfies the following dynamical evolutions:
\bse\label{CDyn}
\begin{align}
&\partial_1\bC_n=\bLd\bC_n+\bC_n\tbLd, \label{CDyna} \\
&\partial_{-1}\bC_n=\bLd^{-1}\bC_n+\bC_n\tbLd^{-1}, \label{CDynb} \\
&\bC_{n+1}(-\tbLd)=\bLd\bC_n. \label{CDync}
\end{align}
\ese
The derivation of these equations is direct. For example, taking the derivative of $\bC_n$ with respect to $x_{-1}$ provides us with
\begin{align*}
\partial_{-1}\bC_n={}&\iint_{D}\rd\zeta(k,k')\bc(k)\partial_{-1}[\rho_n(k)\sigma_n(k')]\tbc(k') \\
={}&\iint_{D}\rd\zeta(k,k')k\bc(k)\rho_n(k)\sigma_n(k')\tbc(k')+\iint_{D}\rd\zeta(k,k')\bc(k)\rho_n(k)\sigma_n(k')\tbc(k')k'.
\end{align*}
And then by making use of the property \eqref{bcDyn}, we end up with \eqref{CDynb}.
Meanwhile, the given Cauchy kernel \eqref{Kernel} implies that $\bOa=-\sum_{i=0}^\infty(-\tbLd)^{-i-1}\bO\bLd^i$, since
\begin{align*}
\tbc(k')\left(-\sum_{i=0}^\infty(-\tbLd)^{-i-1}\bO\bLd^i\right)\bc(k)=-\sum_{i=0}^\infty(-k')^{-i-1}k^i=\frac{1}{k+k'},
\end{align*}
according to \eqref{Omega}. This in turn tells us that $\bOa$ satisfies the algebraic relation
\begin{align}\label{OaDyn}
\bOa\bLd+\tbLd\bOa=\bO.
\end{align}
From equations \eqref{CDyn} and \eqref{OaDyn}, we are able to prove that the infinite matrix $\bU$ obeys the dynamical evolutions as follows:
\bse\label{UDyn}
\begin{align}
&\partial_1\bU_n=\bLd \bU_n+\bU_n\tbLd-\bU_n\bO\bU_n, \label{UDyna} \\
&\partial_{-1}\bU_n=\bLd^{-1}\bU_n+\bU_n\tbLd^{-1}-\bU_n\tbLd^{-1}\bO\bLd^{-1}\bU_n, \label{UDynb} \\
&\bU_{n+1}(-\tbLd)=\bLd\bU_n-\bU_{n+1}\bO\bU_n. \label{UDync}
\end{align}
\ese
We take \eqref{UDynb} as an example to show the derivation. By differentiating equation \eqref{U} with respect to $x_{-1}$, we have
\begin{align*}
\partial_{-1}\bU_n=(1-\bU_n\bOa)(\partial_{-1}\bC_n)-(\partial_{-1}\bU_n)\bOa\bC_n,
\end{align*}
which is equivalent to
\begin{align*}
(\partial_{-1}\bU_n)(1+\bOa\bC_n)=(1-\bU_n\bOa)(\bLd^{-1}\bC_n+\bC_n\tbLd^{-1})=\bLd^{-1}\bC_n+\bU_n\tbLd^{-1}-\bU_n\bOa\bLd^{-1}\bC_n.
\end{align*}
Notice that \eqref{OaDyn} is equivalent to $\bOa\bLd^{-1}+\tbLd^{-1}\bOa=\tbLd^{-1}\bO\bLd^{-1}$.
The above equation is reformulated as
\begin{align*}
(\partial_{-1}\bU_n)(1+\bOa\bC_n)={}&\bLd^{-1}\bC_n+\bU_n\tbLd^{-1}-\bU_n(\tbLd^{-1}\bO\bLd^{-1}-\tbLd^{-1}\bOa)\bC_n \\
={}&\bLd^{-1}\bC_n+\bU_n\tbLd^{-1}(1+\bOa\bC_n)-\bU_n\tbLd^{-1}\bO\bLd^{-1}\bC_n.
\end{align*}
Multiplying this equation by $(1+\bOa\bC_n)^{-1}$ from the right, we end up with \eqref{UDynb} in virtue of \eqref{U}.
Equations \eqref{UDyna} and \eqref{UDync} are proven similarly.
Equations in \eqref{UDyn}, from our view point, form the infinite matrix representation of the \ac{2DTL} of $A_\infty$-type.

Now we present how the scalar closed-form \ac{2DTL} \eqref{A:NL} arises from \eqref{UDyn}.
Taking the derivative of $\ln\tau_n$ gives rise to
\bse\label{tauDyn}
\begin{align*}
\partial_1\ln\tau_n={}&\partial_1\ln[\det(1+\bOa\bC_n)]=\partial_1\tr[\ln(1+\bOa\bC_n)]=\tr[\partial_1\ln(1+\bOa\bC_n)]=\tr[(1+\bOa\bC_n)^{-1}\bOa(\partial_1\bC_n)] \\
={}&\tr[(1+\bOa\bC_n)^{-1}\bOa(\bLd\bC_n+\bC_n\tbLd)]=\tr[(1+\bOa\bC_n)^{-1}\bOa\bLd\bC_n+(1+\bOa\bC_n)^{-1}\bOa\bC_n\tbLd] \\
={}&\tr[\bC_n(1+\bOa\bC_n)^{-1}\bOa\bLd+\bC_n(1+\bOa\bC_n)^{-1}\tbLd\bOa]=\tr(\bU_n\bO)=\tr(\bO\bU_n),
\end{align*}
where the properties \eqref{OaDyn} and \eqref{Commute} are used. Therefore, we have
\begin{align}\label{tauDyna}
\partial_1\ln\tau_n=\bU_n^{(0,0)}, \quad \hbox{and similarly}, \quad \partial_{-1}\ln\tau_n=\bU_n^{(-1,-1)}.
\end{align}
Simultaneously, by performing the shift operation on the tau function with respect to the discrete variable $n$, we obtain
\begin{align*}
\tau_{n+1}={}&\det(1+\bOa\bC_{n+1})=\det\left(1+\bOa\bLd\bC_n(-\tbLd)^{-1}\right)=\det\left(1-(\bO-\tbLd\bOa)\bC_n\tbLd^{-1}\right)
=\det\left(1-\tbLd^{-1}\bO\bC_n+\bOa\bC_n\right) \\
={}&\det(1+\bOa\bC_n)\det\left(1-(1+\bOa\bC_n)^{-1}\tbLd^{-1}\bO\bC_n\right)
=\tau_n\left(1-(\bU_n\tbLd^{-1})^{(0,0)}\right)=\tau_n\left(1-\bU_n^{(0,-1)}\right),
\end{align*}
in which we have made use of \eqref{OaDyn} and \eqref{Rank1}; in other words, the tau function satisfies
\begin{align}\label{tauDynb}
\frac{\tau_{n+1}}{\tau_n}=1-\bU_n^{(0,-1)}, \quad \hbox{and similarly}, \quad \frac{\tau_{n-1}}{\tau_n}=1-\bU_n^{(-1,0)}.
\end{align}
Equations \eqref{tauDyna} and \eqref{tauDynb} transfer the dynamical evolutions of the tau function
with respect to the independent arguments to the entries of the infinite matrix $\bU_n$.
\ese
Notice that
\begin{align*}
\partial_1\partial_{-1}\ln\tau_n=\partial_{-1}(\partial_1\ln\tau_n)=\partial_{-1}\bU_n^{(0,0)},
\end{align*}
because of \eqref{tauDyna}. By taking the $(0,0)$-entry of \eqref{UDynb}, we obtain
\begin{align*}
\partial_1\partial_{-1}\ln\tau_n=\bU_n^{(0,-1)}+\bU_n^{(-1,0)}-\bU_n^{(0,-1)}\bU_n^{(-1,0)}=1-\left(1-\bU_n^{(0,-1)}\right)\left(1-\bU_n^{(-1,0)}\right).
\end{align*}
Therefore, a closed-form equation
\begin{align*}
\partial_{1}\partial_{-1}\ln\tau_n=1-\frac{\tau_{n+1}}{\tau_n}\frac{\tau_{n-1}}{\tau_n}
\end{align*}
is derived, if we replace $\bU_n^{(-1,0)}$ and $\bU_n^{(0,-1)}$ by the tau function according to \eqref{tauDynb}, which can alternatively be written as
\begin{align}\label{A:BL}
\frac{1}{2}\rD_1\rD_{-1}\tau_n\cdot\tau_n=\tau_n^2-\tau_{n+1}\tau_{n-1}
\end{align}
with the help of the bilinear identity $2\partial_1\partial_{-1}\ln\tau_n=\rD_1\rD_{-1}\tau_n\cdot\tau_n/\tau_n^2$,
where the notations $\rD_1$ and $\rD_{-1}$ stand for Hirota's bilinear derivatives\footnote{
For two arbitrary differentiable functions $f(x_1,x_{-1})$ and $g(x_1,x_{-1})$, Hirota's bilinear derivative $\rD$ is defined through
\begin{align*}
\rD_1^i\rD_{-1}^j f\cdot g=\left(\frac{\partial}{\partial_{x_1}}-\frac{\partial}{\partial_{x'_1}}\right)^i
\left(\frac{\partial}{\partial_{x_{-1}}}-\frac{\partial}{\partial_{x'_{-1}}}\right)^j
f(x_1,x_{-1})g(x'_1,x'_{-1})\big|_{x'_1=x_1,x'_{-1}=x_{-1}},
\end{align*}
where $i$ and $j$ are positive integers.
} (see e.g. \cite{Hir04}) with respect to the corresponding arguments $x_1$ and $x_{-1}$, respectively. Equation \eqref{A:BL} is the so-called bilinear \ac{2DTL}, see e.g. \cite{Hir04} and also \cite{JM83,UT84}.
If we introduce a new quantity $\varphi_n\doteq\ln(\tau_{n+1}/\tau_n)=\ln\left(1-\bU_n^{(0,-1)}\right)$, equation \eqref{A:BL} turns out to be its nonlinear form,
which is nothing but the nonlinear \ac{2DTL} of $A_\infty$-type, i.e. \eqref{A:NL}.
We can further write down its Cartan matrix form based on the quantity $\theta_n\doteq\varphi_{n-1}-\varphi_n$,
and subsequently obtain \eqref{A:Cartan}.

\subsection{Lax pair and its adjoint}

The linear equations in \eqref{Linear} not only play a key role in constructing exact solutions of the \ac{2D} Toda equation, but also help to derive Lax pairs of the \ac{2DTL}.
We first concentrate on equation \eqref{u}, i.e. the linear integral equation of $\bu_n(k)$.
Parallel to the equations in \eqref{UDyn}, the wave function $\bu_n(k)$ satisfies the following dynamical relations:
\bse\label{uDyn}
\begin{align}
&\partial_1\bu_n(k)=\bLd\bu_n(k)-\bU_n\bO\bu_n(k), \label{uDyna} \\
&\partial_{-1}\bu_n(k)=\bLd^{-1}\bu_n(k)-\bU_n\tbLd^{-1}\bO\bLd^{-1}\bu_n(k), \label{uDynb} \\
&\bu_{n+1}(k)=\bLd\bu_n(k)-\bU_{n+1}\bO\bu_n(k). \label{uDync}
\end{align}
\ese
The derivation of these equations is straightforward. For instance, taking the derivative of \eqref{u} with respect to $x_{-1}$, we obtain
\begin{align*}
\partial_{-1}\bu_n(k)={}&-(\partial_{-1}\bU_n)\bOa\bc(k)\rho(k)+(1-\bU_n\bOa)\bc(k)[\partial_{-1}\rho_n(k)] \\
={}&-(\bLd^{-1}\bU_n+\bU_n\tbLd^{-1}-\bU_n\tbLd^{-1}\bO\bLd^{-1}\bU_n)\bOa\bc(k)\rho(k)+(1-\bU_n\bOa)\bLd^{-1}\bc(k)\rho_n(k) \\
={}&\bLd^{-1}(1-\bU_n\bOa)\bc(k)\rho_n(k)-\bU_n(\bOa\bLd^{-1}+\tbLd^{-1}\bOa)\bc(k)\rho_n(k)+\bU_n\tbLd^{-1}\bO\bLd^{-1}\bU_n\bOa\bc(k)\rho_n(k).
\end{align*}
Notice that $\bOa\bLd^{-1}+\tbLd^{-1}\bOa=\tbLd^{-1}\bO\bLd^{-1}$. This equation is rewritten as
\begin{align*}
\partial_{-1}\bu_n(k)=\bLd^{-1}(1-\bU_n\bOa)\bc(k)\rho_n(k)-\bU_n\tbLd^{-1}\bO\bLd^{-1}(1-\bU_n\bOa)\bc(k)\rho_n(k),
\end{align*}
which is nothing but \eqref{uDynb}, according to equation \eqref{u}.
Equations listed in \eqref{uDyn} form the infinite vector representation of a Lax pair of the \ac{2DTL}.
To derive a closed-form scalar Lax pair, we need to eliminate $\bLd\bu_n(k)$ in \eqref{uDyn},
and as a consequence, the following relations are obtained:
\begin{align*}
&\partial_1\bu_n(k)=(\bU_{n+1}-\bU_n)\bO\bu_n(k)+\bu_{n+1}(k), \\
&\partial_{-1}\bu_n(k)=\bu_{n-1}(k)-\bLd^{-1}\bU_n\bO\bu_{n-1}(k)-\bU_n\tbLd^{-1}\bO\bu_{n-1}(k)+\bU_n\tbLd^{-1}\bO\bLd^{-1}\bU_n\bO\bu_{n-1}(k).
\end{align*}
The respective $0$th components of the above two equations provide us with
\begin{align*}
\partial_1\phi_n=\left(\bU_{n+1}^{(0,0)}-\bU_n^{(0,0)}\right)\phi_n+\phi_{n+1} \quad \hbox{and} \quad
\partial_{-1}\phi_n=\left(1-\bU_n^{(0,-1)}\right)\left(1-\bU_n^{(-1,0)}\right)\phi_{n-1},
\end{align*}
where $\phi_n\doteq[\bu_n(k)]^{(0)}$. We therefore end up with
\begin{align*}
\partial_1\phi_n=(\partial_1\ln\tau_{n+1}-\partial_1\ln\tau_n)\phi_n+\phi_{n+1}=(\partial_1\varphi_n)\phi_n+\phi_{n+1} \quad \hbox{and} \quad
\partial_{-1}\phi_n=\frac{\tau_{n+1}\tau_{n-1}}{\tau_n^2}\phi_{n-1}=\re^{\varphi_n-\varphi_{n-1}}\phi_{n-1},
\end{align*}
by making use of \eqref{tauDyn}.
By introducing the infinite vector $\Phi={}^{t\!}(\cdots,\phi_{-1},\boxed{\phi_0},\phi_1\cdots)$,
we can alternatively express these two linear equations as a multi-component system composed of
\begin{align}\label{A:Lax}
\partial_1\Phi=\bP\Phi \quad \hbox{and} \quad \partial_{-1}\Phi=\bQ\Phi,
\end{align}
in which
\begin{align*}
\bP=
\begin{pmatrix}
\ddots & \ddots \\
& \partial_1\varphi_{-1} & 1 \\
& & \boxed{\partial_1\varphi_0} & 1 \\
& & & \partial_1\varphi_1 & 1 \\
& & & & \ddots & \ddots
\end{pmatrix}
\quad \hbox{and} \quad
\bQ=
\begin{pmatrix}
\ddots & \ddots \\
& \re^{-\theta_{-1}} & 0 \\
& & \re^{-\theta_0} & \boxed{0} \\
& & & \re^{-\theta_1} & 0 \\
& & & & \ddots & \ddots
\end{pmatrix}.
\end{align*}
We refer to \eqref{A:Lax} as the Lax pair of the \ac{2DTL} of $A_\infty$-type (see \cite{FG80,FG83,UT84}),
since equation \eqref{A:NL} arises as the compatibility condition of the equations in \eqref{A:Lax},
i.e. $\partial_1\partial_{-1}\Phi_n=\partial_{-1}\partial_1\Phi_n$.
Another Lax pair of \eqref{A:NL} in our scheme is derived from the adjoint linear integral equation \eqref{Linearb}.
Performing a similar calculation on \eqref{v}, we can show that the adjoint wave function $\tbv_n(k')$ satisfies the following dynamical relations:
\bse\label{vDyn}
\begin{align}
&\partial_1\tbv_n({k'})=\tbv_n({k'})\tbLd-\tbv_n({k'})\bO\bU_n, \label{vDyna} \\
&\partial_{-1}\tbv_n({k'})=\tbv_n(k')\tbLd^{-1}-\tbv_n(k')\tbLd^{-1}\bO\bLd^{-1}\bU_n, \label{vDynb} \\
&\tbv_{n-1}(k')=-\tbv_n(k')\tbLd+\tbv_n(k')\bO\bU_{n-1}. \label{vDync}
\end{align}
\ese
Following the same idea of deriving \eqref{A:Lax}, we first eliminate $\tbv_n(k')\tbLd^{-1}$ in \eqref{vDyn}.
And then we can find that the quantity $\psi_n\doteq[\tbv_n(k')]^{(0)}$ satisfies
\begin{align*}
\partial_1\psi_n=(\partial_1\ln\tau_{n-1}-\partial_1\ln\tau_n)\psi_n-\psi_{n-1}=-(\partial_1\varphi_{n-1})\psi_n-\psi_{n-1} \quad \hbox{and} \quad
\partial_{-1}\psi_n=-\frac{\tau_{n+1}\tau_{n-1}}{\tau_n^2}\psi_{n+1}=-\re^{\varphi_n-\varphi_{n-1}}\psi_{n+1}.
\end{align*}
We can reformulate these two equations as
\begin{align}\label{A:AdLax}
\partial_1\Psi=\bP'\Psi \quad \hbox{and} \quad \partial_{-1}\Psi=\bQ'\Psi,
\end{align}
with $\Psi={}^{t\!}(\cdots,\psi_{-1},\boxed{\psi_0},\psi_{-1}\cdots)$, where
\begin{align*}
\bP'=
\begin{pmatrix}
\ddots & \ddots \\
& -1 & -\partial_1\varphi_{-2} \\
& & -1 & \boxed{-\partial_1\varphi_{-1}} \\
& & & -1 & -\partial_1\varphi_0 \\
& & & & \ddots & \ddots
\end{pmatrix}
\quad \hbox{and} \quad
\bQ'=
\begin{pmatrix}
\ddots & \ddots \\
& 0 & -\re^{-\theta_{-1}} \\
& & \boxed{0} & -\re^{-\theta_0} \\
& & &  0 & -\re^{-\theta_1} \\
& & & & \ddots & \ddots
\end{pmatrix}.
\end{align*}
The linear problem \eqref{A:AdLax} is referred to as the adjoint Lax pair of the \ac{2DTL} (cf. \cite{NW97}),
as the \ac{2DTL} equation \eqref{A:NL} also arises as the compatibility condition $\partial_1\partial_{-1}\Psi_n=\partial_{-1}\partial_1\Psi_n$ of \eqref{A:AdLax}.

We conclude the main results of this section as theorems below.
\begin{theorem}\label{T:NL}
Consider a linear integral equation of the form
\begin{align}\label{A:Integral}
\bu_n(k)+\iint_D\rd\zeta(l,l')\rho_n(k)\Oa(k,l')\sigma_n(l')\bu_n(l)=\rho_n(k)\bc(k)
\end{align}
with arbitrary measure $\rd\zeta(k,k')$ and integration domain $D$,
where $\rho_n(k)$, $\sigma_n(k')$ and $\Oa(k,k')$ are defined as \eqref{PWF} and \eqref{Kernel}, respectively.
For any solution $\bu_n(k)$ of \eqref{A:Integral},
the quantity
\begin{align}\label{A:DLS}
\varphi_n\doteq\ln\left(1-\bU_n^{(0,-1)}\right)
\end{align}
provides the direct linearising type solution to the nonlinear \ac{2D} Toda equation \eqref{A:NL}, where
\begin{align}\label{A:Potential}
\bU_n\doteq\iint_D\rd\zeta(k,k')\bu_n(k)\tbc(k')\sigma_n(k').
\end{align}
Meanwhile, the wave function $\phi_n\doteq[\bu_n(k)]^{(0)}$ satisfies the linear problem \eqref{A:Lax}.
\end{theorem}
\begin{theorem}\label{T:AdNL}
Consider the adjoint linear integral equation
\begin{align}\label{A:Adjoint}
\tbv_n(k')+\iint_D\rd\zeta(l,l')\rho_n(l)\Oa(l,k')\sigma_n(k')\tbv_n(l')=\tbc(k')\sigma_n(k').
\end{align}
For any solution $\tbv_n(k')$ of \eqref{A:Adjoint},
the quantity $\varphi_n$ defined as \eqref{A:DLS} also provides a solution to the nonlinear equation \eqref{A:NL},
where $\bU_n$ is defined by
\begin{align*}
\bU_n\doteq\iint_D\rd\zeta(k,k')\rho_n(k)\bc(k)\tbv_n(k').
\end{align*}
Furthermore, the adjoint wave function $\psi_n\doteq[\tbv_n(k')]^{(0)}$ provides a solution to the adjoint linear problem \eqref{A:AdLax}.
\end{theorem}
\begin{theorem}\label{T:BL}
The formal tau function
\begin{align}\label{A:tau}
\tau_n\doteq\det(1+\bOa\bC_n)
\end{align}
with $\bOa\doteq-\sum_{i=0}^\infty(-\tbLd)^{-i-1}\bO\bLd^i$ and $\bC_n\doteq\iint_D\rd\zeta(k,k')\rho_n(k)\bc(k)\tbc(k')\sigma_n(k')$
provides the direct linearising type solution to the bilinear \ac{2D} Toda equation \eqref{A:BL}.
\end{theorem}
We remark that the formal tau function is fully determined by the key ingredients of the linear integral equations \eqref{A:Integral} and \eqref{A:Adjoint}, including the plane wave factors, the kernel as well as the integration measure, through $\bOa$ and $\bC_n$,
which provides the most general solution within the \ac{DL} scheme.

The theorems illustrate the \ac{DL} scheme for the \ac{2D} Toda equation of $A_\infty$-type, namely the most general case.
In the coming sections \ref{S:BC} and \ref{S:Reduc}, by performing various reductions on the integration measure $\rd\zeta(k,k')$ and the integration domain $D$,
\ac{2DTL}s associated with $B_\infty$ and $C_\infty$, as well as $A_{r}^{(1)}$, $A_{2r}^{(2)}$, $C_r^{(1)}$ and $D_{r+1}^{(2)}$ are recovered as special reductions,
from the view point of the linear integral equations.

\section{2DTLs associated with $B_\infty$ and $C_\infty$}\label{S:BC}

\subsection{$B_\infty$-type}

To recover the $B_\infty$-type \ac{2DTL}, we identify that the measure in the linear integral equations \eqref{A:Integral} and \eqref{A:Adjoint} satisfies
\begin{align}\label{B:Measure}
\rd\zeta(k,k')=\rd\zeta'(k,k')k, \quad \hbox{in which} \quad \rd\zeta'(k,k')=-\rd\zeta'(k',k), \quad \forall (k,k')\in D,
\end{align}
with symmetric $D$. Such a restriction results in the fact that
\begin{align*}
\tbC_n&=\iint_D\rd\zeta'(k,k')k\bc(k')\tbc(k)\left(-\frac{k}{k'} \right)^n\re^{(k+k')x_{1}+(k^{-1}+{k'}^{-1})x_{-1}} \\
&=\iint_D\rd\zeta'(k',k)k'\bc(k')\tbc(k)\left(-\frac{k'}{k}\right)^{-1-n}\re^{(k+k')x_{1}+(k^{-1}+{k'}^{-1})x_{-1}}=\bC_{-1-n}.
\end{align*}
Notice that $\tbOa=\bOa$ holds identically. By taking the transpose of \eqref{U} and \eqref{tau}, we can derive
\begin{align}\label{B:Reduc}
\tbU_n=\bU_{-1-n} \quad \hbox{and} \quad \tau_n=\tau_{-1-n}, \quad \hbox{and as a result we have} \quad \varphi_n=-\varphi_{-n-2} \quad \hbox{and} \quad \theta_n=\theta_{-1-n}.
\end{align}

The bilinear equation \eqref{A:BL} in this case then reduces to
\begin{align}
&\frac{1}{2}\rD_1\rD_{-1}\tau_0\cdot\tau_0=\tau_0^2-\tau_1\tau_0, \nonumber \\
&\frac{1}{2}\rD_1\rD_{-1}\tau_n\cdot\tau_n=\tau_n^2-\tau_{n+1}\tau_{n-1}, \quad n=1,2,\cdots,
\end{align}
due to \eqref{B:Reduc}, and subsequently, it leads to a reduced nonlinear equation of \eqref{A:NL} in the form of
\begin{align}\label{B:NL}
&\partial_1\partial_{-1}\varphi_0=\re^{\varphi_0}-\re^{\varphi_{1}-\varphi_0}, \nonumber \\
&\partial_1\partial_{-1}\varphi_n=\re^{\varphi_n-\varphi_{n-1}}-\re^{\varphi_{n+1}-\varphi_n},\quad n=1,2,\cdots,
\end{align}
which we refer to as the \ac{2DTL} of $B_\infty$-type. Equation \eqref{B:NL} can also be written as its matrix form
\begin{align}\label{B:Cartan}
\partial_1\partial_{-1}\left(
\begin{array}{c}
\theta_0+\ln2 \\
\theta_1 \\
\theta_2  \\
\vdots
\end{array}
\right)=-
\begin{pmatrix}
2 & -1 \\
-2 & 2 & -1 \\
& -1 & 2 & -1 \\
& & \ddots & \ddots & \ddots
\end{pmatrix}
\left(
\begin{array}{c}
\re^{-(\theta_{0}+\ln2)}\\
\re^{-\theta_1} \\
\re^{-\theta_2} \\
\vdots
\end{array}
\right).
\end{align}
We can easily see that the coefficient matrix is exactly the Cartan matrix associated with $B_\infty$.

The Lax pair and the adjoint one of \eqref{B:NL} still take the same form of \eqref{A:Lax} and \eqref{A:AdLax}, respectively,
subject to the additional conditions for $\varphi_n$ and $\theta_n$ given in \eqref{B:Reduc}.
To be more precise, the linear problem takes the same form of \eqref{A:Lax}, but with the Lax matrices given by
\begin{align*}
\bP=
\begin{pmatrix}
\ddots & \ddots \\
&-\partial_1\varphi_0 & 1 \\
& & 0 & 1 & & &\\
& & & \boxed{\partial_1\varphi_0} & 1 \\
& & & & \partial_1\varphi_1 & 1 \\
& & & & & \partial_1\varphi_2 & 1 \\
& & & & & & \ddots & \ddots
\end{pmatrix}
\quad \hbox{and} \quad
\bQ=
\begin{pmatrix}
\ddots & \ddots & \\
& \re^{-\theta_{1}} & 0 \\
& & \re^{-\theta_{0}}& 0 \\
& & & \re^{-\theta_0}& \boxed{0} \\
& & & & \re^{-\theta_1}&  0 \\
& & & & & \re^{-\theta_2}& 0 \\
& & & & & & \ddots & \ddots
\end{pmatrix},
\end{align*}
respectively. Similarly, the adjoint linear problem takes the form of \eqref{A:AdLax}, in which the Lax matrices $\bP'$ and $\bQ'$ are given as follows:
\begin{align*}
\bP'=
\begin{pmatrix}
\ddots & \ddots \\
& -1 & \partial_1\varphi_{1} \\
& & -1 & \partial_1\varphi_0 \\
& & & -1 & \boxed{0} \\
& & & & -1 & -\partial_1\varphi_0 \\
& & & & & -1 & -\partial_1\varphi_1 \\
& & & & & & \ddots &\ddots
\end{pmatrix},
\quad
\bQ'=
\begin{pmatrix}
\ddots& \ddots & & & & & \\
& 0 & -\re^{-\theta_1}  \\
& & 0 &-\re^{-\theta_{0}} \\
& & & \boxed{0} & -\re^{-\theta_0} \\
& & & & 0 & -\re^{-\theta_1} \\
& & & & & 0 & -\re^{-\theta_2} \\
& & & & & & \ddots &\ddots
\end{pmatrix}.
\end{align*}

\subsection{$C_\infty$-type}
For the \ac{2D} Toda equation of $C_\infty$-type, we consider a symmetric integration domain $D$ and simultaneously impose the symmetry condition
\begin{align}\label{C:Measure}
\rd\zeta(k,k')=\rd\zeta(k',k), \quad \forall (k,k')\in D.
\end{align}
Taking the transpose of $\bC_n$, we observe that
\begin{align*}
\tbC_n&=\iint_D\rd\zeta(k,k')\bc(k')\tbc(k)\left(-\frac{k}{k'}\right)^n\re^{(k+k')x_{1}+(k^{-1}+{k'}^{-1})x_{-1}} =\iint_D\rd\zeta(k',k)\bc(k')\tbc(k)\left(-\frac{k'}{k}\right)^{-n}\re^{(k+k')x_{1}+(k^{-1}+{k'}^{-1})x_{-1}}=\bC_{-n},
\end{align*}
and subsequently $\tbU_n=\bU_{-n}$. Again, observing that $\tbOa=\bOa$ and simultaneously making use of $\tbC_n=\bC_{-n}$, we obtain
\begin{align}\label{C:Reduc}
\tau_{n}=\tau_{-n}, \quad \hbox{and therefore}, \quad \varphi_{n}=-\varphi_{-n-1}, \quad \hbox{and} \quad \theta_n=\theta_{-n}.
\end{align}

By performing the reduction conditions \eqref{C:Reduc} on the \ac{2DTL} of $A_\infty$-type,
the bilinear equation \eqref{A:BL} and the nonlinear equation \eqref{A:NL} reduce to multi-component systems
\begin{align}\label{C:BL}
&\frac{1}{2}\rD_1\rD_{-1}\tau_0\cdot\tau_0=\tau_0^2-\tau_1^2, \nonumber \\
&\frac{1}{2}\rD_1\rD_{-1}\tau_n\cdot\tau_n=\tau_n^2-\tau_{n+1}\tau_{n-1},\quad n=1,2,\cdots
\end{align}
and
\begin{align}\label{C:NL}
&\partial_1\partial_{-1}\varphi_0=\re^{2\varphi_0}-\re^{\varphi_1-\varphi_0}, \nonumber \\
&\partial_1\partial_{-1}\varphi_n=\re^{\varphi_n-\varphi_{n-1}}-\re^{\varphi_{n+1}-\varphi_n},\quad n=1,2,\cdots,
\end{align}
respectively, in which the latter is referred to as the \ac{2DTL} of $C_\infty$-type.
Equation \eqref{C:NL} can alternatively be written in its the matrix form by the quantity $\theta_n$ as follows:
\begin{align}\label{C:Cartan}
\partial_1\partial_{-1}\left(
\begin{array}{c}
\theta_0 \\
\theta_1 \\
\theta_2 \\
\vdots
\end{array}
\right)=-
\begin{pmatrix}
2 & -2 \\
-1 & 2 & -1 \\
& -1 & 2 & -1 \\
& & \ddots & \ddots &\ddots
\end{pmatrix}
\left(
\begin{array}{c}
\re^{-\theta_{0}} \\
\re^{-\theta_1} \\
\re^{-\theta_2} \\
\vdots
\end{array}
\right),
\end{align}
with the entries of the Cartan matrix of $C_\infty$ appearing as the coefficients.

The Lax pair and its adjoint of \eqref{C:NL} also follow from \eqref{A:Lax} and \eqref{A:AdLax}, respectively, subject to \eqref{C:Reduc}.
In other words, the linear problem and the adjoint linear problem are still written in their respective forms \eqref{A:Lax} and \eqref{A:AdLax};
however, the Lax matrices in the $C_\infty$ class are given by
\begin{align*}
\bP=
\begin{pmatrix}
&\ddots& \ddots \\
& & -\partial_1\varphi_1 & 1 \\
& & & -\partial_1\varphi_0 & 1 \\
& & & & \boxed{\partial_1\varphi_0} & 1 \\
& & & & & \partial_1\varphi_1 & 1 \\
& & & & & & \partial_1\varphi_2 & 1 \\
& & & & & & & \ddots &\ddots
\end{pmatrix},
\quad
\bQ=
\begin{pmatrix}
\ddots & \ddots \\
& \re^{-\theta_{2}} & 0 \\
& & \re^{-\theta_{1}} & 0 \\
& & & \re^{-\theta_0} & \boxed{0} \\
& & & & \re^{-\theta_1} & 0 \\
& & & & & \re^{-\theta_1} & 0 \\
& & & & & & \ddots & \ddots
\end{pmatrix},
\end{align*}
and
\begin{align*}
\bP'=
\begin{pmatrix}
\ddots &\ddots & \\
& -1 & \partial_1\varphi_2 \\
& & -1 & \partial_1\varphi_1 \\
& & & -1 & \boxed{\partial_1\varphi_0} \\
& & & & -1& -\partial_1\varphi_0 \\
&& & & & -1& -\partial_1\varphi_1 \\
& & & & & & \ddots &\ddots &
\end{pmatrix},
\quad
\bQ'=
\begin{pmatrix}
& \ddots & \ddots \\
& & 0 & -\re^{-\theta_{2}} \\
& & & 0 & -\re^{-\theta_{1}} \\
& & & & \boxed{0} & -\re^{-\theta_0} \\
& & & & & 0 & -\re^{-\theta_1} \\
& & & & & & 0 & -\re^{-\theta_2} \\
& & & & & & & \ddots & \ddots
\end{pmatrix},
\end{align*}
respectively.

The results in this section show that theorems \ref{T:NL}, \ref{T:AdNL} and \ref{T:BL} are also applicable to the \ac{2DTL}s of $B_\infty$- and $C_\infty$-types,
subject to the reductions that the arbitrary integration measure and domain in the theorems obey \eqref{B:Measure} and \eqref{C:Measure}, respectively.
To put it another way, we have successfully constructed the direct linearising type solutions to the $B_\infty$-type and $C_\infty$-type \ac{2D} Toda equations.

\section{2DTLs associated with $A_{r}^{(1)}$, $A_{2r}^{(2)}$, $C_r^{(1)}$ and $D_{r+1}^{(2)}$}\label{S:Reduc}

We have discussed the \ac{2DTL}s of $A_\infty$-, $B_\infty$- and $C_\infty$-types from the linear integral equations.
In these cases, there are effectively two spectral parameters, namely $k$ and $k'$, playing roles in the solution structure, leading to (2+1)-dimensional integrable models.
In this section, we consider the so-called periodic reduction on the integration measure,
by restricting the spectral parameters $k$ and $k'$ on the curve $k^\cN=(-k')^\cN$ for integer $\cN$,
namely there is effectively only one spectral parameter.
As a result, the \ac{2DTL}s of $A_{r}^{(1)}$-, $A_{2r}^{(2)}$-, $C_r^{(1)}$- and $D_{r+1}^{(2)}$-types are recovered,
as (1+1)-dimensional integrable systems.

\subsection{$A_{r}^{(1)}$-type}

We perform the periodicity condition $k^{r+1}=(-k')^{r+1}$ on the $A_\infty$-type Toda equation. This is realised by taking a particular measure in the form of
\begin{align}\label{Ar1:Measure}
\rd\zeta(k,k')=\sum_{j=1}^{\varphiup(r+1)}\frac{1}{2\pi\ri}\frac{\rd\lambda_j(k)\rd k'}{k'+\oa_{r+1}^{(j)}k}
\end{align}
and a particular domain $D=\Gamma\times\Gamma'$ with $\Gamma$ being an arbitrary integration domain for $k$ and $\Gamma'$ being a closed contour in terms of $k'$ that containing singularities $-\oa_{r+1}^{(j)}k$.
Here $\oa_{\cN}^{(j)}$ stand for the primitive $\cN$th roots of unity and $\varphiup(\cdot)$ denotes Euler's totient function.
For example, when $\cN=4$ we have $\varphiup(\cN)=2$; therefore, there are two primitive $4$th roots of unity, given by $\oa_4^{(1)}=\oa_4$ and $\oa_4^{(2)}=\oa_4^3$, respectively, where $\oa_4=\exp(2\pi\ri/4)$.

By Cauchy's residue theorem, the linear integral equations \eqref{A:Integral} and \eqref{A:Adjoint} under such a constraint hence turn out to be
\begin{align}\label{Ar1:Integral}
\bu_n(k)+\sum_{j=1}^{\varphiup(r+1)}\int_{\Gamma_j}\rd\ld_j(l)\rho_n(k)\Oa({k,-\oa_{r+1}^{(j)} l})\sigma_n({-\oa_{r+1}^{(j)} l})\bu_n(l)=\rho_n(k)\bc(k)
\end{align}
and
\begin{align}\label{Ar1:Adjoint}
\tbv_n(k')+\sum_{j=1}^{\varphiup(r+1)}\int_{\Gamma_j}\rd\ld_j(l)\rho_n(l)\Oa(l,k')\sigma_n(k')\tbv_n(-\oa_{r+1}^{(j)} l)=\tbc(k')\sigma_n(k'),
\end{align}
namely linear integral equations only involve single integrations in terms of the spectral parameter $k$,
in which $\Gamma_j$ and $\rd\ld_j(l)$ are arbitrary integration domains and measures reduced from $D$ and $\rd\zeta(k,k')$, respectively.
We remark that at this stage we still do not fix the integration domains and measures, as this shall bring us the most general solution space for the reduced equations. Once explicit forms are selected, special classes of solutions can be constructed.
For example, when the integration domains are chosen to be closed contours that contain a finite number of higher-order poles,
the generalised Cauchy matrix solutions can be obtained, see section \ref{S:Sol} for more detail.

Simultaneously, the reduced measure \eqref{Ar1:Measure} also leads to
\begin{align}\label{Ar1:Potential}
\bU_n=\sum_{j=1}^{\varphiup(r+1)}\int_{\Gamma_j}\rd\ld_j(k)\bu_n(k)\tbc(-\oa_{r+1}^{(j)} k)\sigma_n(-\oa_{r+1}^{(j)} k),
\end{align}
as well as
\begin{align}\label{Ar1:Dispersion}
\bC_n=\sum_{j=1}^{\varphiup(r+1)}\int_{\Gamma_j}\rd\ld_j(k)\rho_n(k)\bc(k)\tbc({-\oa_{r+1}^{(j)} k})\sigma_n({-\oa_{r+1}^{(j)} k}).
\end{align}
Notice that
\begin{align*}
\rho_{n+r+1}(k)\sigma_{n+r+1}(-\oa_{r+1}^{(j)} k)
=\re^{(k-\oa_{r+1}^{(j)}k)x_1+(k^{-1}-(\oa_{r+1}^{(j)}k)^{-1})x_{-1}}\left(\frac{1}{\oa_{r+1}^{(j)}}\right)^{n+r+1}
=\rho_{n}(k)\sigma_{n}(-\oa_{r+1}^{(j)} k).
\end{align*}
It is easily verified from \eqref{Ar1:Dispersion} that $\bC_{n+r+1}=\bC_n$. Subsequently,
\begin{align}\label{Ar1:Reduc}
\bU_{n+r+1}=\bU_n \quad \hbox{and} \quad \tau_{n+r+1}=\tau_n, \quad \hbox{and futhermore we have} \quad
\varphi_{n+r+1}=\varphi_n \quad \hbox{and} \quad \theta_{n+r+1}=\theta_n.
\end{align}

The periodicity of $\tau_n$ then reduces \eqref{A:BL} to a multi-component system of (1+1)-dimensional bilinear equations
\begin{align}\label{Ar1:BL}
&\frac{1}{2}\rD_1\rD_{-1}\tau_0\cdot\tau_0=\tau_0^2-\tau_1\tau_{r}, \nonumber \\
&\frac{1}{2}\rD_1\rD_{-1}\tau_n\cdot\tau_n=\tau_n^2-\tau_{n+1}\tau_{n-1},\quad n=1,2,\cdots,r-1, \\
&\frac{1}{2}\rD_1\rD_{-1}\tau_{r}\cdot\tau_{r}=\tau_{r}^2-\tau_0\tau_{r-1}, \nonumber
\end{align}
which is the bilinear form of the $A_r^{(1)}$-type \ac{2D} Toda system. The nonlinear \ac{2DTL} of $A_r^{(1)}$-type reads
\begin{align}\label{Ar1:NL}
&\partial_1\partial_{-1}\varphi_0=\re^{\varphi_0-\varphi_r}-\re^{\varphi_1-\varphi_0}, \nonumber \\
&\partial_1\partial_{-1}\varphi_n=\re^{\varphi_n-\varphi_{n-1}}-\re^{\varphi_{n+1}-\varphi_n},\quad n=1,2,\cdots,r-1, \\
&\partial_1\partial_{-1}\varphi_{r}=\re^{\varphi_{r}-\varphi_{r-1}}-\re^{\varphi_0-\varphi_{r}}, \nonumber
\end{align}
as it follows from \eqref{A:NL} in virtue of the periodicity of $\varphi_n$ given in \eqref{Ar1:Reduc}.
Equation \eqref{Ar1:NL} can alternatively be written as
\begin{align}
\partial_1\partial_{-1}
\left(
\begin{array}{c}
\theta_0 \\
\theta_1 \\
\vdots \\
\theta_{r-1} \\
\theta_{r}
\end{array}
\right)
=-
\begin{pmatrix}
2 & -1 & & & -1 \\
-1 & 2 & -1 & & \\
& \ddots & \ddots& \ddots & \\
& & -1 & 2 & -1\\
-1 & & & -1 & 2
\end{pmatrix}
\left(
\begin{array}{c}
\re^{-\theta_0} \\
\re^{-\theta_1} \\
\vdots \\
\re^{-\theta_{r-1}} \\
\re^{-\theta_{r}}
\end{array}
\right),
\end{align}
where the coefficient matrix is exactly the same as the Cartan matrix for $A_r^{(1)}$.
In general, the nonlinear equations for the $A_r^{(1)}$ case can only be expressed as multi-component systems like \eqref{Ar1:NL}.
However, the $A_1^{(1)}$ case is an exception.
Due to $\tau_{n+2}=\tau_n$, we have $\varphi_1=\ln(\tau_2/\tau_1)=\ln(\tau_{0}/\tau_{1})=-\varphi_{0}$,
and therefore, the first equation in \eqref{Ar1:NL} is closed as a scalar equation, taking the form of \eqref{sG},
i.e. the sinh-Gordon equation.

The Lax pair and the adjoint Lax pair of equation \eqref{Ar1:NL} are also derived from the measure reduction \eqref{Ar1:Measure}.
Notice that $\rho_{n+r+1}(k)=k^{r+1}\rho_n(k)$. Shifting $n$ by $r+1$ units in \eqref{u}, we obtain
\begin{align*}
\bu_{n+r+1}(k)=(1-\bU_{n+r+1}\bOa)\bc(k)\rho_{n+r+1}(k)=k^{r+1}(1-\bU_n\bOa)\bc(k)\rho_n(k).
\end{align*}
In other words, the wave function $\phi_n$ obeys the quasi-periodicity
\begin{align*}
\quad\phi_{n+r+1}=k^{r+1}\phi_n, \quad \hbox{and similarly}, \quad \psi_{n+r+1}=k^{-{(r+1)}}\psi_n,
\end{align*}
where we have replaced $k'$ by $k$ in the equation for $\psi_n$, since \eqref{Ar1:Reduc} implies that $k^{r+1}=(-k')^{r+1}$.
Hence, the Lax pair \eqref{A:Lax} reduces to
\begin{align}\label{Ar1:Lax}
\partial_1\Phi=\bP\Phi, \quad \partial_{-1}\Phi=\bQ\Phi,
\end{align}
where $\Phi={}^{t\!}(\phi_0,\phi_1,\cdots,\phi_r)$, the matrices $\bP$ and $\bQ$ are given by
\begin{align*}
\bP=
\begin{pmatrix}
\partial_1\varphi_0 & 1 \\
& \partial_1\varphi_1 & 1 \\
& & \ddots& \ddots \\
& & & \partial_1\varphi_{r-1} & 1\\
k^{r+1} & & & & \partial_1\varphi_{r}
\end{pmatrix}
\quad \hbox{and} \quad
\bQ=
\begin{pmatrix}
0 & & & & k^{-(r+1)}\re^{-\theta_0} \\
\re^{-\theta_1} & 0 \\
& \ddots & \ddots \\
& & \re^{-\theta_{r-1}} & 0 \\
& & & \re^{-\theta_{r}} & 0
\end{pmatrix},
\end{align*}
respectively; the adjoint Lax pair \eqref{A:AdLax} reduces to
\begin{align}\label{Ar1:AdLax}
\partial_1\Psi=\bP'\Psi, \quad \partial_{-1}\Psi=\bQ'\Psi,
\end{align}
where $\Psi={}^{t\!}(\psi_0,\psi_1,\cdots,\psi_r)$, and $\bP'$ and $\bQ'$ are defined as
\begin{align*}
\bP'=
\begin{pmatrix}
-\partial_1\varphi_0 & & & & -k^{r+1}\\
-1& -\partial_1\varphi_1 \\
& \ddots& \ddots \\
& & -1 & -\partial_1\varphi_{r-1} \\
& & & -1 & -\partial_1\varphi_{r}
\end{pmatrix}
\quad \hbox{and} \quad
\bQ'=
\begin{pmatrix}
0 & -\re^{-\theta_0} \\
& 0 &-\re^{-\theta_1} \\
& & \ddots & \ddots \\
& & & 0 & -\re^{-\theta_{r-1}} \\
-k^{-(r+1)}\re^{-\theta_{r}} & & & & 0
\end{pmatrix},
\end{align*}
respectively.

\subsection{$A_{2r}^{(2)}$-type}

We now take a special measure
\begin{align}\label{BN:Measure}
\rd\zeta(k,k')=k\sum_{j=1}^{\varphiup(\cN)}\frac{1}{2\pi\ri}\left(\frac{\rd\lambda_j(k)\rd k'}{k'+\oa_{\cN}^{(j)}k}-\frac{\rd k\rd\lambda_j(k')}{k+\oa_{\cN}^{(j)}k'}\right).
\end{align}
This leads to the reduced linear integral equation
\begin{align}\label{BN:Integral}
\bu_n(k)+\sum_{j=1}^{\varphiup(\cN)}\int_{\Gamma_j}\rd\lambda_j(l)\left[\rho_n(k)\Oa{(k,-\oa_{\cN}^{(j)}l)}\sigma_n(-\oa_{\cN}^{(j)}l)l\bu_n(l)
-\rho_n(k)\Oa(k,l)\sigma_n(l)(-\oa_{\cN}^{(j)}l)\bu_n({-\oa_{\cN}^{(j)}l})\right]=\rho_n(k)\bc(k),
\end{align}
and its adjoint equation
\begin{align}\label{BN:Adjoint}
\tbv_n(k')+\sum_{j=1}^{\varphiup(\cN)}\int_{\Gamma_j}\rd\lambda_j(l)\left[\rho_n(l)\Oa(l,k')\sigma_n(k')l\,\tbv_n(-\oa_{\cN}^{(j)}l)
-\rho_n(-\oa_{\cN}^{(j)}l)\Oa(-\oa_{\cN}^{(j)}l,k')\sigma_n(k')(-\oa_{\cN}^{(j)}l)\tbv_n({l})\right]
=\tbc(k')\sigma_n(k').
\end{align}
Likewise, the infinite matrices $\bU_n$ and $\bC_n$ under such a reduction become
\begin{align}\label{BN:Potential}
\bU_n=\sum_{j=1}^{\varphiup(\cN)}\int_{\Gamma_j}\rd\lambda_j(k)\left[k\bu_n(k)\tbc(-\oa_{\cN}^{(j)}k)\sigma_n(-\oa_{\cN}^{(j)}k)-(-\oa_{\cN}^{(j)}k)\bu_n(-\oa_{\cN}^{(j)} k)\tbc(k)\sigma_n(k)\right]
\end{align}
and
\begin{align}\label{BN:Dispersion}
\bC_n=\sum_{j=1}^{\varphiup(\cN)}\int_{\Gamma_j}\rd\lambda_j(k)\left[k\rho_n(k)\bc(k)\tbc(-\oa_{\cN}^{(j)}k)\sigma_n(-\oa_{\cN}^{(j)}k)
-(-\oa_{\cN}^{(j)}k)\rho_n(-\oa_{\cN}^{(j)}k)\bc(-\oa_{\cN}^{(j)}k)\tbc(k)\sigma_n(k)\right],
\end{align}
respectively.

For the $A_{2r}^{(2)}$ case, we set $\cN=2r+1$. Straightforward computation shows that $\tbC_n=\bC_{-1-n}$, as well as $\bC_{n+2r+1}=\bC_n$, which imply that $\bU_{n+2r+1}=\bU_n$ and $\tbU_n=\bU_{-1-n}$.
Based on these relations, we obtain the reductions on $\tau_n$, $\varphi_n$ as well as $\theta_n$ as follows:
\begin{align}\label{A2r2:Reduc}
\tau_{n+2r+1}=\tau_n, \quad \tau_n=\tau_{-1-n}, \quad \varphi_{n+2r+1}=\varphi_n, \quad \varphi_n=-\varphi_{-2-n}, \quad \theta_{n+2r+1}=\theta_n, \quad \theta_n=\theta_{-1-n}.
\end{align}

Performing the reduction of $\tau_n$ on \eqref{A:BL}, we immediately obtain the bilinear system
\begin{align}\label{A2r2:BL}
&\frac{1}{2}\rD_1\rD_{-1}\tau_0\cdot\tau_0=\tau_0^2-\tau_1\tau_0, \nonumber \\
&\frac{1}{2}\rD_1\rD_{-1}\tau_n\cdot\tau_n=\tau_n^2-\tau_{n+1}\tau_{n-1},\quad n=1,2,\cdots,r-1, \\
&\frac{1}{2}\rD_1\rD_{-1}\tau_{r}\cdot\tau_{r}=\tau_{r}^2-\tau_{r-1}^2. \nonumber
\end{align}
Similarly, the reduction on $\varphi_n$ provides us with the \ac{2DTL} of $A_{2r}^{(2)}$-type, i.e.
\begin{align}\label{A2r2:NL}
&\partial_1\partial_{-1}\varphi_0=\re^{\varphi_0}-\re^{\varphi_{1}-\varphi_0}, \nonumber \\
&\partial_1\partial_{-1}\varphi_n=\re^{\varphi_n-\varphi_{n-1}}-\re^{\varphi_{n+1}-\varphi_n},\quad n=1,2,\cdots,r-2, \\
&\partial_1\partial_{-1}\varphi_{r-1}=\re^{\varphi_{r-1}-\varphi_{r-2}}-\re^{-2\varphi_{r-1}}, \nonumber
\end{align}
and its Cartan matrix form is given by 	
\begin{align}\label{A2r2:Cartan}
\partial_1\partial_{-1}
\left(
\begin{array}{c}
\theta_0+\ln 2 \\
\theta_1 \\
\vdots \\
\theta_{r-1} \\
\theta_{r}
\end{array}
\right)
=-
\begin{pmatrix}
2 & -1 \\
-2 & 2 & -1 \\
& \ddots & \ddots & \ddots \\
& & -1 & 2 & -1 \\
& & & -2 & 2
\end{pmatrix}
\left(
\begin{array}{c}
\re^{-(\theta_0+\ln2)} \\
\re^{-\theta_1} \\
\vdots \\
\re^{-\theta_{r-1}} \\
\re^{-\theta_{r}}
\end{array}
\right).
\end{align}
When $r=1$, equation \eqref{A2r2:NL} turns out to be a closed-form scalar equation due to $\varphi_1=-\varphi_0$,
which is the nothing but the Tzitzeica equation \eqref{Tz1}.

Observing that the wave functions $\phi_n$ and $\psi_n$ still satisfy the periodicity conditions
$\phi_{n+2r+1}=k^{2r+1}\phi_n$ and $\psi_{n+2r+1}=k^{-(2r+1)}\psi_n$, we obtain the associated linear problems for equation \eqref{A2r2:NL},
including the linear problem
\begin{align}\label{A2r2:Lax}
\partial_1\Phi=\bP\Phi, \quad \partial_{-1}\Phi=\bQ\Phi
\end{align}
with $\Phi={}^{t\!}(\phi_0,\phi_1,\cdots,\phi_{2r})$ and Lax matrices
\begin{align*}
\bP=
\begin{pmatrix}
\partial_1\varphi_0 & 1 \\
& \ddots& \ddots \\
& & \partial_1\varphi_{r-1} & 1 \\
& & & -\partial_1\varphi_{r-1} & 1 \\
& & & & \ddots& \ddots \\
& & & & & -\partial_1\varphi_0 & 1 \\
k^{2r+1} & & & & & & 0
\end{pmatrix}
\quad \hbox{and} \quad
\bQ=
\begin{pmatrix}
0 & & & & & & k^{-(2r+1)}\re^{-\theta_0} \\
\re^{-\theta_1}& 0 & & & & & & \\
& \ddots & \ddots& & & & &\\
& & \re^{-\theta_{r}}& 0  & & & &\\
& & & \re^{-\theta_{r-1}} & 0 & &\\
& & & & \ddots & \ddots& &  \\
& & & & & \re^{-\theta_0} & 0
\end{pmatrix},
\end{align*}
as well as the adjoint linear problem
\begin{align}\label{A2r2:AdLax}
\partial_1\Psi=\bP'\Psi, \quad \partial_{-1}\Psi=\bQ'\Psi,
\end{align}
in which $\Psi={}^{t\!}(\psi_0,\psi_1,\cdots,\psi_{2r})$, and
\begin{align*}
\bP'=
\begin{pmatrix}
0 & & & & & & -k^{2r+1}\\
-1 & -\partial_1\varphi_0 \\
& \ddots & \ddots& & & & \\
& & -1& -\partial_1\varphi_{r-1} \\
& & &  -1 & \partial_1\varphi_{r-1} \\
& & & & \ddots & \ddots & \\
& & & & & -1 & \partial_1\varphi_0
\end{pmatrix},
\quad
\bQ'=
\begin{pmatrix}
0 & -\re^{-\theta_0} \\
& \ddots & \ddots & & & & \\
& & 0 & -\re^{-\theta_{r}} \\
& & & 0 & -\re^{-\theta_{r-1}} \\
& & & & \ddots & \ddots & \\
& & & & & 0 & -\re^{-\theta_1} \\
-k^{-(2r+1)}\re^{-\theta_0} & & & & & & 0
\end{pmatrix}.
\end{align*}

We comment that the measure \eqref{BN:Measure} in a sense is a combination of \eqref{B:Measure} and \eqref{Ar1:Measure},
as it leads to a reduction which simultaneously satisfies the anti-symmetry and periodicity conditions.
However, taking such a measure is not the only reduction to the \ac{2DTL} of $A_{2r}^{(2)}$-type.
In fact, we can alternatively consider the measure
\begin{align}\label{CN:Measure}
\rd\zeta(k,k')=\sum_{j=1}^{\varphiup(\cN)}\frac{1}{2\pi\ri}\left(\frac{\rd\lambda_j(k)\rd k'}{k'+\oa_{\cN}^{(j)}k}+\frac{\rd k\,\rd\lambda_j(k')}{k+\oa_{\cN}^{(j)}k'}\right),
\end{align}
which should be thought of as imposing \eqref{C:Measure} and \eqref{Ar1:Measure} simultaneously.
This brings us the linear integral equation
\begin{align}\label{CN:Integral}
\bu_n(k)+\sum_{j=1}^{\varphiup(\cN)}\int_{\Gamma_j}\rd\lambda_j(l)\left[\rho_n(k)\Oa{(k,-\oa_{\cN}^{(j)} l)}\sigma_n(-\oa_{\cN}^{(j)} l)\bu_n(l)+\rho_n(k)\Oa(k,l)\sigma_n(l)\bu_n({-\oa_{\cN}^{(j)}l})\right]=\rho_n(k)\bc(k),
\end{align}
and the adjoint linear integral equation
\begin{align}\label{CN:Adjoint}
\tbv_n(k')+\sum_{j=1}^{\varphiup(\cN)}\int_{\Gamma_j}\rd\lambda_j(l)\left[\rho_n(l)\Oa(l,k')\sigma_n(k')\tbv_n(-\oa_{\cN}^{(j)}l)
+\rho_n(-\oa_{\cN}^{(j)}l)\Oa(-\oa_{\cN}^{(j)}l,k')\sigma_n(k')\tbv_n({l})\right]=\bc(k')\sigma_n(k'),
\end{align}
as well as the infinite matrices
\begin{align}\label{CN:Potential}
\bU_n=\sum_{j=1}^{\varphiup(\cN)}\int_{\Gamma_j}\rd\lambda_j(k)\left[\bu_n(k)\tbc(-\oa_{\cN}^{(j)}k)\sigma_n(-\oa_{\cN}^{(j)}k)
+\bu_n(-\oa_{\cN}^{(j)}k)\tbc(k)\sigma_n(k)\right]
\end{align}
and
\begin{align}\label{CN:Dispersion}
\bC_n=\sum_{j=1}^{\varphiup(\cN)}\int_{\Gamma_j}\rd\lambda_j(k)\left[\rho(k)\bc(k)\tbc(-\oa_{\cN}^{(j)}k)\sigma(-\oa_{\cN}^{(j)}k)
+\rho(-\oa_{\cN}^{(j)}k)\bc(-\oa_{\cN}^{(j)}k)\tbc(k)\sigma(k)\right].
\end{align}
Therefore, by setting $\cN=2r+1$ we can follow the same procedure and find reduction conditions
\begin{align*}
\tau_{n+2r+1}=\tau_n, \quad \tau_n=\tau_{-n}, \quad \varphi_{n+2r+1}=\varphi_{n}, \quad \varphi_n=-\varphi_{-n-1}, \quad \theta_{n+2r+1}=\theta_n, \quad \theta_n=\theta_{-n}
\end{align*}
as well as
\begin{align*}
\phi_{n+2r+1}=k^{2r+1}\phi_{n} \quad \hbox{and} \quad \psi_{n+2r+1}=k^{-(2r+1)}\psi_{n}
\end{align*}
on the \ac{2DTL} of $A_\infty$-type, leading to the equivalent results in the $A_{2r}^{(2)}$ class,
compared with \eqref{A2r2:BL}, \eqref{A2r2:NL}, \eqref{A2r2:Cartan}, \eqref{A2r2:Lax} as well as \eqref{A2r2:AdLax}.
For example, in this case, the simplest model expressed by the potential $\varphi_n$ when $r=1$ gives rise to another form of the Tzitzeica equation, namely \eqref{Tz2}.

\subsection{$C_r^{(1)}$-type}

The $C_r^{(1)}$-type \ac{2D} Toda equation follows from the $2r$-periodic reduction of the $C_\infty$ case.
For this reason, we set $\cN=2r$ in \eqref{CN:Measure}, \eqref{CN:Integral}, \eqref{CN:Adjoint}, \eqref{CN:Potential} and \eqref{CN:Dispersion}.
We can directly verify that $\bC_{n+2r}=\bC_n$ and $\tbC_{n}=\bC_{-n}$ and subsequently $\bU_{n+2r}=\bU_n$ and $\tbU_{n}=\bU_{-n}$,
i.e. both quantities satisfy the symmetry and periodicity conditions simultaneously.
Based on these conditions, we conclude that
\begin{align}\label{Cr1:Reduc}
\tau_n=\tau_{n+2r}, \quad \tau_n=\tau_{-n}, \quad \varphi_{n+2r}=\varphi_{n}, \quad \varphi_n=-\varphi_{-n-1}, \quad
\theta_{n+2r}=\theta_n, \quad \hbox{and} \quad \theta_n=\theta_{-n}.
\end{align}

Performing these reductions, we obtain the bilinear system
\begin{align}\label{Cr1:BL}
&\frac{1}{2}\rD_1\rD_{-1}\tau_0\cdot\tau_0=\tau_0^2-\tau_1^2, \nonumber \\
&\frac{1}{2}\rD_1\rD_{-1}\tau_n\cdot\tau_n=\tau_n^2-\tau_{n+1}\tau_{n-1}, \quad n=1,2,\cdots,r-1, \\
&\frac{1}{2}\rD_1\rD_{-1}\tau_r\cdot\tau_r=\tau_r^2-\tau_{r-1}^2, \nonumber
\end{align}
and the nonlinear system
\begin{align}\label{Cr1:NL}
&\partial_1\partial_{-1}\varphi_0=\re^{2\varphi_0}-\re^{\varphi_1-\varphi_0}, \nonumber \\
&\partial_1\partial_{-1}\varphi_n=\re^{\varphi_n-\varphi_{n-1}}-\re^{\varphi_{n+1}-\varphi_n}, \quad n=1,2,\cdots,r-2 \\
&\partial_1\partial_{-1}\varphi_{r-1}=\re^{\varphi_{r-1}-\varphi_{r-2}}-\re^{-2\varphi_{r-1}}, \nonumber
\end{align}
from \eqref{A:BL} and \eqref{A:NL}, respectively, in which the latter is referred to as the $C_r^{(1)}$-type \ac{2DTL}.
Alternatively, we can reformulate \eqref{Cr1:NL} as
\begin{align*}
\partial_1\partial_{-1}\left(
\begin{array}{c}
\theta_0 \\
\theta_1 \\
\vdots \\
\theta_{r-1} \\
\theta_{r}
\end{array}
\right)
=-
\begin{pmatrix}
2 & -2 & & &  \\
-1 & 2 & -1 & &  \\
& \ddots & \ddots& \ddots&  \\
& &  -1 & 2 & -1\\
& &  & -2 & 2
\end{pmatrix}
\left(
\begin{array}{c}
\re^{-\theta_0} \\
\re^{-\theta_1} \\
\vdots \\
\re^{-\theta_{r-1}} \\
\re^{-\theta_{r}}
\end{array}
\right),
\end{align*}
from which we can observe the Cartan matrix of $C_r^{(1)}$.
We comment that the relation $\varphi_{-1}=-\varphi_0$ holds identically for $r=1$,
and in this case equation \eqref{Cr1:NL} turns out to be the sinh--Gordon equation (which has been discussed in the $A_1^{(1)}$ class.
Therefore, we normally start with $r=2$ in this class.

Notice that the periodicity of $\bU_n$ provides us with $\phi_{n+2r}=k^{2r}\phi_n$ and $\psi_{n+2r}=k^{-2r}\psi_n$, where we have made use of $k^{2r}=(-k')^{2r}$.
We obtain from \eqref{A:Lax} and \eqref{A:AdLax} the Lax pair
\begin{align}\label{Cr1:Lax}
\partial_1\Phi=\bP\Phi, \quad \partial_{-1}\Phi=\bQ\Phi
\end{align}
with $\Phi={}^{t\!}(\phi_0,\phi_1,\cdots,\phi_{2r-1})$, where
\begin{align*}
\bP=
\begin{pmatrix}
\partial_1\varphi_0 & 1 \\
& \ddots & \ddots \\
& & \partial_1\varphi_{r-1} & 1 \\
& & & -\partial_1\varphi_{r-1} & 1 \\
& & & & \ddots & \ddots \\
& & & & & -\partial_1\varphi_1 & 1 \\
k^{2r} & & & & & & -\partial_1\varphi_0
\end{pmatrix},
\quad
\bQ=
\begin{pmatrix}
0 & & & & & & k^{-2r}\re^{-\theta_0} \\
\re^{-\theta_1} & 0 \\
& \ddots & \ddots \\
& & \re^{-\theta_{r}} & 0 \\
& & & \re^{-\theta_{r-1}} & 0 \\
& & & & \ddots & \ddots \\
& & & & & \re^{-\theta_1} & 0
\end{pmatrix},
\end{align*}
as well as the adjoint Lax pair
\begin{align}\label{Cr1:AdLax}
\partial_1\Psi=\bP'\Psi, \quad \partial_{-1}\Psi=\bQ'\Psi,
\end{align}
with $\Psi={}^{t\!}(\psi_0,\psi_1,\cdots,\psi_{2r-1})$, in which
\begin{align*}
\bP'=
\begin{pmatrix}
\partial_1\varphi_0 & & & & & & -k^{2r}\\
-1& -\partial_1\varphi_0 \\
&  \ddots & \ddots \\
& & -1& -\partial_1\varphi_{r-1} \\
& & &  -1 & \partial_1\varphi_{r-1} \\
& & & & \ddots& \ddots \\
& & & & & -1 & \partial_1\varphi_1
\end{pmatrix},
\quad
\bQ'=
\begin{pmatrix}
0 & -\re^{-\theta_0}\\
& \ddots & \ddots \\
& & 0 &-\re^{-\theta_r} \\
& & & 0 &-\re^{-\theta_{r-1}} \\
& & & & \ddots & \ddots \\
& & & & & 0 &-\re^{-\theta_2} \\
-k^{-2r}\re^{-\theta_{1}} & & & & & & 0
\end{pmatrix},
\end{align*}
respectively, as the two linear problems for the \ac{2DTL} of $C_\infty$-type.

\subsection{$D_{r+1}^{(2)}$-type}

Finally, we discuss the $D_{r+1}^{(2)}$-type \ac{2DTL}.
This is done by taking $\cN=2r+2$ in \eqref{BN:Measure}, \eqref{BN:Integral}, \eqref{BN:Adjoint}, \eqref{BN:Potential} and \eqref{BN:Dispersion},
which results in the fact that in the $D_{r+1}^{(2)}$ class the reduction conditions are as follows:
\begin{align}\label{Dr2:Reduc}
\tau_n=\tau_{n+2r+2}, \quad \tau_n=\tau_{-1-n}, \quad \varphi_n=\varphi_{n+2r+2}, \quad \varphi_n=-\varphi_{-2-n}, \quad
\theta_n=\theta_{n+2r+2}, \quad \theta_n=\theta_{-1-n}.
\end{align}
Making use of the constraints in \eqref{Dr2:Reduc}, we obtain the reduced bilinear system
\begin{align}\label{Dr2:BL}
&\frac{1}{2}\rD_1\rD_{-1}\tau_0\cdot\tau_0=\tau_0^2-\tau_1\tau_0, \nonumber \\
&\frac{1}{2}\rD_1\rD_{-1}\tau_n\cdot\tau_n=\tau_n^2-\tau_{n+1}\tau_{n-1},\quad n=1,2, \cdots,r-1, \\
&\frac{1}{2}\rD_1\rD_{-1}\tau_r\cdot\tau_r=\tau_r^2-\tau_r\tau_{r-1}, \nonumber
\end{align}
the nonlinear system
\begin{align}\label{Dr2:NL}
&\partial_1\partial_{-1}\varphi_0=\re^{\varphi_0}-\re^{\varphi_{1}-\varphi_0}, \nonumber \\
&\partial_1\partial_{-1}\varphi_n=\re^{\varphi_n-\varphi_{n-1}}-\re^{\varphi_{n+1}-\varphi_n},\quad n=1,2,\cdots,r-2, \\
&\partial_1\partial_{-1}\varphi_{r-1}=\re^{\varphi_{r-1}-\varphi_{r-2}}-\re^{-\varphi_{r-1}}, \nonumber
\end{align}
as well as the Cartan matrix form as follows:
\begin{align}\label{Dr2:Cartan}
\partial_1\partial_{-1}
\left(
\begin{array}{c}
\theta_0+\ln2 \\
\theta_1 \\
\vdots \\
\theta_{r-1} \\
\theta_r+\ln2
\end{array}
\right)
=-
\begin{pmatrix}
2 & -1 \\
-2 & 2 & -1 \\
& \ddots & \ddots & \ddots & \\
& & -1 & 2 & -2 \\
& & & -1 & 2
\end{pmatrix}
\left(
\begin{array}{c}
\re^{-(\theta_0+\ln2)} \\
\re^{-\theta_1} \\
\vdots \\
\re^{-\theta_{r-1}} \\
\re^{-(\theta_r+\ln2)}
\end{array}
\right).
\end{align}
The $r=1$ case in this class also leads to the $A_1^{(1)}$ class.
This is because when $r=1$, we have $\varphi_1=-\varphi_1$, i.e. $\varphi_1=0$, according to \eqref{Dr2:Reduc}.
Thus, we obtain from \eqref{Dr2:NL} a closed-form scalar equation
\begin{align*}
\partial_1\partial_{-1}\varphi_0=\re^{\varphi_0}-\re^{-\varphi_0},
\end{align*}
which is exactly the same as the sinh--Gordon equation \eqref{sG}, up to a scaling transformation.
In other words, we have to start with $D_3^{(2)}$ in order to obtain nontrivial equations.

The linear problems for \eqref{Dr2:NL} are derived accordingly, with the help of the quasi-periodicity conditions
$\phi_{n+2r+2}=k^{2r+2}\phi_n$ and $\psi_{n+2r+2}=k^{-(2r+2)}\psi_n$.
The Lax pair takes the form of
\begin{align}
\partial_1\Phi=\bP\Phi, \quad \partial_{-1}\Phi=\bQ\Phi,
\end{align}
for $\Phi={}^{t\!}(\phi_0,\cdots,\phi_{2r+1})$, where the Lax matrices $\bP$ and $\bQ$ are as follows:
\begin{align*}
\bP=
\begin{pmatrix}
\partial_1\varphi_0 & 1 \\
& & \ddots & \ddots\\
& & & \partial_1\varphi_{r-1} & 1 \\
& & & & 0 & 1 \\
& & & & & -\partial_1\varphi_{r-1} & 1 \\
& & & & & & \ddots & \ddots \\
& & & & & & & -\partial_1\varphi_0 & 1\\
k^{2r+2} & & & & & & & & 0
\end{pmatrix},
\quad
\bQ=
\begin{pmatrix}
0 & & & & & & k^{-(2r+2)}\re^{-\theta_0} \\
\re^{-\theta_1}& 0 \\
& \ddots & \ddots \\
& & \re^{-\theta_r} & 0 \\
& & & \re^{-\theta_r} & 0 \\
& & & & \ddots & \ddots \\
& & & & & \re^{-\theta_0} & 0
\end{pmatrix}.
\end{align*}
The adjoint Lax pair is composed of the linear equations
\begin{align}
\partial_1\Psi=\bP'\Psi \quad \hbox{and} \quad \partial_{-1}\Psi=\bQ'\Psi,
\end{align}
where $\Psi={}^{t\!}(\psi_0,\cdots,\psi_{2r+1})$, and the infinite matrices $\bP'$ and $\bQ'$ take their respective forms of
\begin{align*}
\bP'=
\begin{pmatrix}
0 & & & & & & & -k^{2r+2} \\
-1 & -\partial_1\varphi_0 \\
& \ddots & \ddots \\
& & -1 & -\partial_1\varphi_{r-1} \\
& & & -1 & 0 \\
& & & & -1 & \partial_1\varphi_{r-1} \\
& & & & & \ddots & \ddots \\
& & & & & & -1 & \partial_1\varphi_0
\end{pmatrix}
\quad \hbox{and} \quad
\bQ'=
\begin{pmatrix}
0 & -\re^{-\theta_0} \\
& \ddots & \ddots \\
& & 0 & -\re^{-\theta_r} \\
& & & 0 & -\re^{-\theta_r} \\
& & & & \ddots & \ddots \\
& & & & & 0 & -\re^{-\theta_1} \\
-k^{-(2r+2)}\re^{-\theta_0} & & & & & & 0
\end{pmatrix}.
\end{align*}

We have established the \ac{DL} scheme for the \ac{2D} Toda-type equations of
$A_{r+1}^{(1)}$-, $A_{2r}^{(2)}$-, $C_r^{(1)}$- and $D_{r+1}^{(2)}$-types for positive integers $r$,
by identifying the corresponding linear integral equations for each class of nonlinear equations.
The main results in the theorems in section \ref{S:A} are inherited for these classes of \ac{2DTL}s,
subject to their respective measure degenerations, leading to the direct linearising type solutions.

\section{General formulae of Cauchy matrix type solutions}\label{S:Sol}

We now discuss a special class of solutions called Cauchy matrix solutions for all the \ac{2D} Toda-type systems.
These solutions are constructed by taking particular measures containing an arbitrary number of singularities of arbitrary order,
as special cases of the direct linearising type solutions presented in the previous sections.

In this section, our aim is to propose a general formula of the Cauchy matrix solution for each class of the \ac{2DTL}.
Notice that the potential for the \ac{2D} Toda-type equations has a common form
\begin{align}\label{2DTL:Sol}
\varphi_n\doteq\ln\left(1-\bU_n^{(0,-1)}\right),
\end{align}
with $\rho_n(k)$ and $\sigma_n(k')$ defined as \eqref{PWF}. Hence, below we only present the general formula of $\bU_n^{(\bar{i},\bar{j})}$
for arbitrary integers $\bar{i}$ and $\bar{j}$, class by class.
For convenience, we omit the suffix $n$ in our formulae, if not necessary.

\subsection{$A_\infty$-type}

We consider the measure
\begin{align}\label{A:Pole}
\mathrm{d}\zeta(k,k')=\sum_{j=1}^{N}\sum_{j'=1}^{N'}A_{j,j'}(s_j-1)!(s'_{j'}-1)!\frac{1}{(2\pi \mathrm{i})^2}\frac{1}{(k-k_j)^{s_j}}\frac{1}{(k'-k'_{j'})^{s'_{j'}}}\mathrm{d}k\mathrm{d}k',
\end{align}
and simultaneously the integration domain $D=\Gamma\times\Gamma'$ in which $\Gamma$ and $\Gamma'$ are the two separate contours on the $k$- and $k'$-planes
containing singularities $k_j$ of order $s_j$ and $k'_j$ of order $s'_{j'}$ inside for $j=1,\cdots,N$ and $j'=1,\cdots,N'$, respectively.
The residue theorem then reduces \eqref{A:Integral} and \eqref{A:Potential} to
\begin{align}\label{A:u1}
\bu(k)+\sum_{j=1}^{N}\sum_{j'=1}^{N'}A_{j,j'}\left(\partial_{k'_{j'}}^{s'_{j'}-1}\frac{\rho(k)\sigma{(k'_{j'})}}{k+k'_{j'}}\right)\left(\partial_{k_j}^{s_j-1}\bu{(k_j)}\right)=\rho(k)\bc(k)
\end{align}
and
\begin{align}\label{A:U1}
\bU=\sum_{j=1}^{N}\sum_{j'=1}^{N'}A_{j,j'}\left(\partial_{k_j}^{s_j-1}\bu({k_j})\right)\left(\partial_{k'_{j'}}^{s'_{j'}-1}\tbc{(k'_{j'})}\sigma{(k'_{j'})}\right),
\end{align}
respectively. For convenience below, we introduce finite matrices
\begin{align*}
\bA=
\begin{pmatrix}
A_{1,1} & \cdots & A_{1,N'} \\
\vdots & & \vdots \\
A_{N,1} & \cdots & A_{N,N'}
\end{pmatrix}
\quad \hbox{and} \quad
\bM=
\begin{pmatrix}
M_{1,1}& \cdots &
M_{1,N} \\
\vdots& & \vdots \\
M_{N',1} & \cdots & M_{N',N}
\end{pmatrix}
\quad \hbox{where} \quad
M_{j',j}=\partial_{k_{j}}^{s_j-1}\partial_{k'_{j'}}^{s'_{j'}-1}\frac{\rho{(k_j)}\sigma{(k'_{j'})}}{k_j+k'_{j'}},
\end{align*}
diagonal block matrices
\begin{align*}
\mathscr{C}=\diag(\mathscr{C}_1,\cdots,\mathscr{C}_{N})
\quad \hbox{and} \quad
\bcscr'=\diag(\tbcscr'_1,\cdots,\tbcscr'_{N'}),
\end{align*}
in which
\begin{align*}
\mathscr{C}_{j}={}^{t\!}\left(\tbc({k_{j}}),\cdots,\frac{\partial_{k_{j}}^{s_{j}-1}\tbc{(k_{j})}}{(s_j-1)!}\right), \quad
\tbcscr'_{{j}}=\left(\frac{\partial_{k'_{j'}}^{s'_{j'}-1}\tbc({k'_{j'}})}{(s'_{j'}-1)!},\cdots,\tbc{(k'_{j'})}\right),
\end{align*}
diagonal matrices
\begin{align*}
\bD=\diag((s_1-1)!,\cdots,(s_N-1)!) \quad \hbox{and} \quad \bD'=\diag((s'_1-1)!,\cdots,(s'_{N'}-1)!),
\end{align*}
as well as finite row and column vectors
\begin{align*}
\tbr=(\tbr_{1},\cdots,\tbr_{N}) \quad \hbox{and} \quad \bs={}^{t\!}(\tbs_{1},\cdots,\tbs_{N'}),
\end{align*}
where
\begin{align*}
\tbr_{j}=\left(\frac{\partial^{s_{j}-1}_{k_{j}}\rho{(k_{j})}}{(s_{j}-1)!},\cdots,\rho{(k_j)}\right)
\quad \hbox{and} \quad
\tbs_{{j'}}=\left(\sigma{(k'_{j'})},\cdots,\frac{\partial^{s'_{j'}-1}_{k'_{j'}}\sigma{(k'_{j'})}}{(s'_{j'}-1)!}\right).
\end{align*}
With these objects, equation \eqref{A:U1} is expressed as
\begin{align}\label{A:U2}
\bU=\tbuscr\bA\bD'\bcscr'\bs, \quad \hbox{where} \quad
\tbuscr=\left(\partial_{k_{1}}^{s_1-1}\bu{(k_1)},\cdots,\partial_{k_{N}}^{s_N-1}\bu{(k_N)}\right).
\end{align}
Meanwhile, from equation \eqref{A:u1} we can write down
\begin{align}\label{A:u2}
\tbuscr+\tbuscr\bA\bM=\tbr\bcscr\bD.
\end{align}
This in turn implies that we are able to eliminate $\tbuscr$ in \eqref{A:U2} through \eqref{A:u2} and obtain
\begin{align*}
\bU=\tbr\mathscr{C}\bD(\bI+\bA\bM)^{-1}\bA\bD'\bcscr'\bs.
\end{align*}
From the first glance, the sizes of the left and right sides of the above equation do not match.
But in fact, the multiplication on the right hand side should be thought of as that of block matrices;
in other words, the right hand side reads
\begin{align*}
\left((s_1-1)!\tbr_1\mathscr{C}_1,\cdots,(s_N-1)!\tbr_N\mathscr{C}_N\right)(1+\bA\bM)^{-1}\bA
\begin{pmatrix}
(s'_1-1)!\tbcscr'_1\bs_{1} \\
\vdots \\
(s'_{N'}-1)!\tbcscr'_{N'}\bs_{N'}
\end{pmatrix}
\end{align*}
which is of size
\begin{align*}
&(\hbox{block row vector with components of size $\infty\times1$})_{1\times N} \\
&\qquad\times(\hbox{finite matrix with entries of size $1\times 1$})_{N\times N'}\\
&\qquad\qquad\times(\hbox{block column vector with components of size $1\times\infty$})_{N'\times 1},
\end{align*}
because the $j$th component in the block row vector takes the form of an $\infty\times 1$ column vector
\begin{align*}
(s_j-1)!\tbr_j\mathscr{C}_j=(s_j-1)!\sum_{i=0}^{s_j-1}\frac{\partial^{s_{j}-1-i}_{k_{j}}\rho{(k_{j})}}{(s_{j}-1-i)!}\frac{\partial_{k_{j}}^{i}\bc{(k_{j})}}{i!},
\end{align*}
and the $j'$th component in the block column vector takes the form of a $1\times\infty$ row vector
\begin{align*}
(s'_{j'}-1)!\tbcscr'_{j'}\bs_{j'}=(s'_{j'}-1)!\sum_{i'=0}^{s'_{j'}-1}\frac{\partial_{k'_{j'}}^{s'_{j'}-1-i'}\tbc({k'_{j'}})}{(s'_{j'}-1-i')!}\frac{\partial^{i'}_{k'_{j'}}\sigma{(k'_{j'})}}{i'!},
\end{align*}
according to the multiplication of block matrices.
Thus, the right hand side of $\bU$ is indeed of size $\infty\times\infty$.
Taking the $(\bar{i},\bar{j})$-entry of $\bU$, we finally reach to the general solution structure
\begin{align}\label{A:Sol}
\bU^{(\bar{i},\bar{j})}=\tbr\bK^{(\bar{i})}\bD(\bI+\bA\bM)^{-1}\bA\bD'\bK'^{(\bar{j})}\bs,
\end{align}
where $\bK^{(\bar{i})}$ and $\bK'^{(\bar{j})}$ are defined as
\begin{align*}
\bK^{(\bar{i})}=\diag\left(\bK_1^{(\bar{i})},\cdots,\bK_N^{(\bar{i})}\right) \quad \hbox{and} \quad
{\bK'}^{(\bar{j})}=\diag\left({}^{t\!}{\bK'}_{1}^{(\bar{j})},\cdots,{}^{t\!}{\bK'}_{N'}^{(\bar{j})}\right),
\end{align*}
respectively, with the blocks given as follows:
\begin{align*}
\bK_j^{(\bar{i})}={}^{t\!}\left(k_j^{\bar{i}},\cdots,\frac{\partial_{k_{j}}^{s_{j}-1}k_j^{\bar{i}}}{(s_j-1)!}\right), \quad
{}^{t\!}{\bK'}_{j'}^{(\bar{j})}=\left(\frac{\partial_{k'_{j'}}^{s'_{j'}-1}{k'}^{\bar{j}}_{j'}}{(s'_{j'}-1)!},\cdots,{k'}_{j'}^{\bar{j}}\right).
\end{align*}

\subsection{$B_\infty$-type}

We respect \eqref{B:Measure} and take a degenerate measure
\begin{align}\label{B:Pole}
\mathrm{d}\zeta(k,k')=\sum_{j,j'=1}^{N}A_{j,j'}(s_j-1)!(s_{j'}-1)!\frac{1}{(2\pi\mathrm{i})^2}\frac{1}{(k-k_j)^{s_j}}\frac{1}{(k'-k_{j'})^{s_{j'}}}k\mathrm{d}k\mathrm{d}k',\quad
\hbox{where} \quad A_{j,j'}=-A_{j',j},
\end{align}
and the integration domain $D=\Gamma\times\Gamma'$ with $\Gamma$ and $\Gamma'$ containing singularities
$k_j$ of order $s_j$ and $k_{j'}$ of order $s_{j'}$ on their respective complex planes, for $j=1,\cdots,N$ and $j=1,\cdots,N$.
The residue theorem reduces the linear integral equation \eqref{A:Integral} and the potential \eqref{A:Potential} to
\begin{align}\label{B:u1}
\bu(k)+\sum_{j,j'=1}^{N}A_{j,j'}\left(\partial_{k_{j'}}^{s_{j'}-1}\frac{\rho(k)\sigma(k_{j'})}{k+k_{j'}}\right)\left(\partial_{k_j}^{s_j-1}k_j\bu(k_j)\right)=\rho(k)\bc(k)
\end{align}
and
\begin{align}\label{B:U1}
\bU=\sum_{j,j'=1}^{N}A_{j,j'}\left(\partial_{k_j}^{s_j-1}k_j\bu(k_j)\right)\left(\partial_{k_{j'}}^{{s}_{j'}-1}\tbc(k_{j'})\sigma(k_{j'})\right),
\end{align}
respectively. For convenience, we introduce an antisymmetric matrix $\bA$ and the Cauchy matrix $\bM$ given by
\begin{align*}
\bA=
\begin{pmatrix}
A_{1,1} & \cdots & A_{1,N} \\
\vdots & & \vdots \\
A_{N,1} & \cdots & A_{N,N}
\end{pmatrix}
\quad \hbox{and} \quad
\bM=
\begin{pmatrix}
M_{1,1} & \cdots &
M_{1,N} \\
\vdots & & \vdots \\
M_{N,1} & \cdots & M_{N,N}
\end{pmatrix},
\end{align*}
respectively, in which
\begin{align*}
A_{j,j'}=-A_{j'j}, \quad M_{j',j}=\partial_{k_{j}}^{s_j-1}\partial_{k_{j'}}^{s_{j'}-1}\frac{k_j\rho{(k_j)}\sigma(k_{j'})}{k_j+k_{j'}},
\end{align*}
as well as a diagonal matrix $D$ and diagonal block matrices as follows:
\begin{align*}
\bD=\diag((s_1-1)!,\cdots,(s_N-1)!), \quad
\mathscr{C}=\mathrm{diag}(\mathscr{C}_1,\cdots,\mathscr{C}_{N}),
\quad
\bcscr'=\mathrm{diag}(\tbcscr'_1,\cdots,\tbcscr'_{N}),
\end{align*}
where the blocks $\mathscr{C}_{{j}}$ and $\tbcscr'_{{j}}$ are defined as follows:
\begin{align*}
\mathscr{C}_{j}={}^{t\!}\left(k_j\tbc(k_{j}),\cdots\frac{\partial_{k_j}^{s_j-1}k_j\tbc(k_j)}{(s_j-1)!}\right), \quad
\tbcscr'_{j}=\left(\frac{\partial_{k_{j'}}^{s_{j'}-1}\tbc(k_{j'})}{(s_{j'}-1)!} ,\cdots,\tbc(k_{j'})\right).
\end{align*}
Simultaneously, we also adopt the notations $\tbr$ and $\bs$ for a row vector and a column vector, defined by
\begin{align*}
\tbr=(\tbr_{1},\cdots,\tbr_{N}) \quad \hbox{and} \quad \bs={}^{t\!}(\tbs_{1},\cdots,\tbs_{N}),
\end{align*}
respectively, where
\begin{align*}
\tbr_{j}=\left(\frac{\partial^{s_{j}-1}_{k_{j}}\rho{(k_{j})}}{(s_{j}-1)!},\cdots,\rho{(k_j)}\right)
\quad \hbox{and} \quad
\tbs_{{j'}}=\left(\sigma{(k_{j'})},\cdots,\frac{\partial^{s_{j'}-1}_{k_{j'}}\sigma{(k_{j'})}}{(s_{j'}-1)!}\right).
\end{align*}
With these new symbols, we can rewrite $\bU$ in \eqref{B:U1} as
\begin{align}\label{B:U2}
\bU={\tbuscr}\bA\bD\bcscr'\bs, \quad \hbox{where} \quad {\tbuscr}=\left(\partial_{k_{1}}^{s_1-1}k_1\bu(k_1),\cdots,\partial_{k_{N}}^{s_N-1}k_N\bu(k_N)\right).
\end{align}
At the same time, we obtain from \eqref{B:u1} the matrix equation
\begin{align}\label{B:u2}
\tbuscr+\tbuscr\bA\bM=\tbr\bcscr\bD,
\end{align}
which together with \eqref{B:U2} provides us with the Cauchy matrix expression of $\bU$ as follows:
\begin{align*}
\bU=\tbr\mathscr{C}\bD(\bI+\bA\bM)^{-1}\bA\bD\bcscr'\bs.
\end{align*}
Then the $(\bar{i},\bar{j})$-entry of $\bU$ takes the expression
\begin{align}
\bU^{(\bar{i},\bar{j})}=\tbr\bK^{(\bar{i})}\bD(\bI+\bA\bM)^{-1}\bA\bD\bK'^{(\bar{j})}\bs.
\end{align}
Here $\bK^{(\bar{i})}$ and ${\bK'}^{(\bar{j})}$ are block diagonal matrices
\begin{align*}
\bK^{(\bar{i})}=\diag\left(\bK_1^{(\bar{i})},\cdots,\bK_N^{(\bar{i})}\right) \quad \hbox{and} \quad
{\bK'}^{(\bar{j})}=\diag\left({}^{t\!}{\bK'}_{1}^{(\bar{j})},\cdots,{}^{t\!}{\bK'}_{N}^{(\bar{j})}\right),
\end{align*}
where the blocks are finite vectors as follows:
\begin{align*}
\bK_j^{(\bar{i})}={}^{t\!}\left(k_j^{\bar{i}+1},\cdots,\frac{\partial_{k_{j}}^{s_{j}-1}k_j^{\bar{i}+1}}{(s_j-1)!}\right), \quad
{}^{t\!}{\bK'}_{j'}^{(\bar{j})}=\left(\frac{\partial_{k_{j'}}^{s_{j'}-1}{k}^{\bar{j}}_{j'}}{(s_{j'}-1)!},\cdots,{k}_{j'}^{\bar{j}}\right).
\end{align*}

\subsection{$C_\infty$-type}

We consider a symmetric measure in the $C_\infty$ class following \eqref{C:Measure} as follows:
\begin{align}\label{C:Pole}
\mathrm{d}\zeta(k,k')=\sum_{j,j'=1}^{N}A_{j,j'}(s_j-1)!(s_{j'}-1)!\frac{1}{(2\pi\mathrm{i})^2}\frac{1}{(k-k_j)^{s_j}}\frac{1}{(k'-k_{j'})^{s_{j'}}}\mathrm{d}k\mathrm{d}k', \quad
\hbox{where} \quad A_{j,j'}=A_{j',j},
\end{align}
and still let the domain $D$ being the tensor of two contours containing the singularities $k_j$ and $k_{j'}$ on their respective complex planes.
By comparing \eqref{C:Pole} with \eqref{A:Pole}, we observe that the only difference is that additional constraints
\begin{align}\label{C:Sol}
A_{j,j'}=A_{j',j}, \quad k'_{j'}=k_{j'}, \quad s'_{j'}=s_{j'} \quad \hbox{and} \quad N'=N
\end{align}
are imposed in the $C_\infty$ case. Therefore, equation \eqref{A:Sol} subject to \eqref{C:Sol} provides the general formula
for the Cauchy matrix solution of the \ac{2DTL} of $C_\infty$-type.

\subsection{$A_{r}^{(1)}$-type}

Recall that the linear integral equations and the potential of $A_{r}^{(1)}$-type are given by \eqref{Ar1:Integral} and \eqref{Ar1:Potential}.
To construct the Cauchy matrix solution, we take a special measure in the form of
\begin{align}\label{Ar1:Pole}
\mathrm{d}\lambda_j(k)=\sum_{j'=1}^{N_j}A_{j,j'}(s_{j,j'}-1)!\frac{1}{2\pi\mathrm{i}}\frac{1}{(k-k_{j,j'})^{s_{j,j'}}}\rd k,
\end{align}
and let $k_{j,j'}$ be the singularities of order $s_{j,j'}$ surrounded by their respective contours $\Gamma_j$.
And then the residue theorem implies that the linear integral equation \eqref{Ar1:Integral} reduces to
\begin{align}\label{Ar1:u1}
\bu(k)+\sum_{j=1}^{\varphiup(r+1)}\sum_{j'=1}^{N_j}A_{j,j'}\frac{\mathrm{d}^{s_{j,j'}-1}}{\mathrm{d}k_{j,j'}^{s_{j,j'}-1}}
\left(\frac{\rho(k)\sigma(-\oa_{r+1}^{(j)}k_{j,j'})}{k-\oa_{r+1}^{(j)}k_{j,j'}}\bu(k_{j,j'})\right)=\rho(k)\bc(k).
\end{align}
Simultaneously, the potential $\bU$ under \eqref{Ar1:Pole} turns out to be
\begin{align}\label{Ar1:U1}
\bU=\sum_{j=1}^{\varphiup(r+1)}\sum_{j'=1}^{N_j}A_{j,j'}\frac{\mathrm{d}^{s_{j,j'}-1}}{\mathrm{d}k_{j,j'}^{s_{j,j'}-1}}
\left(\bu(k_{j,j'})\tbc(-\oa_{r+1}^{(j)}k_{j,j'})\sigma(-\oa_{r+1}^{(j)}k_{j,j'})\right).
\end{align}
For convenience, we introduce block vectors
\begin{align*}
\tbuscr=\left(\tbuscr_{1,1},\cdots,\tbuscr_{1,N_1};\cdots;\tbuscr_{{\varphiup(r+1)},1},\cdots,\tbuscr_{\varphiup(r+1),N_{\varphiup(r+1)}}\right), \quad \hbox{in which} \quad
\tbuscr_{j,j'}=\left(\frac{\partial_{k_{j,j'}}^{s_{j,j'}-1}\bu(k_{j,j'})}{(s_{j,j'}-1)!},\cdots,\bu(k_{j,j'})\right),
\end{align*}
and
\begin{align}
\tbr=\left(\tbr_{1,1},\cdots,\tbr_{1,N_1};\cdots;\tbr_{{\varphiup(r+1)},1},\cdots,\tbr_{\varphiup(r+1),N_{\varphiup(r+1)}}\right), \quad \hbox{with} \quad
\tbr_{j,j'}=\left(\frac{\partial_{k_{j,j'}}^{s_{j,j'}-1}\rho(k_{j,j'})}{(s_{j,j'}-1)!},\cdots,\rho(k_{j,j'})\right),
\end{align}
as well as
\begin{align}
\bs={}^{t\!}\left(\tbs_{1,1},\cdots,\tbs_{1,N_1};\cdots;\tbs_{{\varphiup(r+1)},1},\cdots,\tbs_{\varphiup(r+1),N_{\varphiup(r+1)}}\right), \quad \hbox{with} \quad
\tbs_{j,j'}=\left(\sigma(-\oa_{r+1}^{(j)}k_{j,j'}),\cdots,\frac{\partial_{k_{j,j'}}^{s_{j,j'}-1}\sigma(-\oa_{r+1}^{(j)}k_{j,j'})}{(s_{j,j'}-1)!}\right),
\end{align}
and also block diagonal matrices
\begin{align*}
\bA=\mathrm{diag}\left(\bA_{1,1},\cdots,\bA_{1,N_1};\cdots;\bA_{\varphiup(r+1),1},\cdots,\bA_{\varphiup(r+1),N_{\varphiup(r+1)}}\right), \quad \hbox{in which} \quad
\bA_{j,j'}=A_{j,j'}\bI_{s_{j,j'}\times s_{j,j'}},
\end{align*}
and
\begin{align*}
\bcscr=\mathrm{diag}\left(\bcscr_{1,1},\cdots,\bcscr_{1,N_1};\cdots;\bcscr_{\varphiup(r+1),1},\cdots,\bcscr_{\varphiup(r+1),N_{\varphiup(r+1)}}\right) \quad \hbox{and} \quad
\bcscr'=\mathrm{diag}\left(\bcscr'_{1,1},\cdots,\bcscr'_{1,N_1};\cdots;\bcscr'_{\varphiup(r+1),1},\cdots,\bcscr'_{\varphiup(r+1),N_{\varphiup(r+1)}}\right),
\end{align*}
where
\begin{align*}
\bcscr_{j,j'}=
\begin{pmatrix}
\bc(k_{j,j'}) \\
\vdots & \ddots & \\
\frac{\partial_{k_{j,j'}}^{s_{j,j'}-1}\bc(k_{j,j'})}{(s_{j,j'}-1)!} & \cdots & \bc(k_{j,j'})
\end{pmatrix}
\quad \hbox{and} \quad
\bcscr'_{j,j'}=
\begin{pmatrix}
\tbc(-\oa_{r+1}^{(j)}k_{j,j'})& & \\
\vdots& \ddots& \\
\frac{\partial_{k_{j,j'}}^{s_{j,j'}-1}\tbc(-\oa_{r+1}^{(j)}k_{j,j'})}{(s_{j,j'}-1)!} & \cdots & \tbc(-\oa_{r+1}^{(j)}k_{j,j'})
\end{pmatrix},
\end{align*}
as well as
\begin{align*}
\bD=\left(\bD_{1,1},\cdots,\bD_{1,N_1};\cdots;\bD_{\varphiup(r+1),1},\cdots,\bD_{\varphiup(r+1),N_{\varphiup(r+1)}}\right), \quad \hbox{with} \quad
\bD_{j,j'}=(s_{j,j'}-1)!\bI_{s_{j,j'}\times s_{j,j'}},
\end{align*}
and finally a Cauchy matrix
\begin{align*}
\bM=
\begin{pmatrix}
M_{1,1;1,1} & \cdots & M_{1,1;1,N_1} & \cdots & M_{1,1;\varphiup(r+1),1} & \cdots & M_{1,1;\varphiup(r+1),N_{\varphiup(r+1)}} \\
\vdots &  & \vdots & & \vdots &  & \vdots  \\
M_{1,N_1;1,1} & \cdots & M_{1,N_1;1,N_1} & \cdots & M_{1,N_1;\varphiup(r+1),1} & \cdots & M_{1,N_1;\varphiup(r+1),N_{\varphiup(r+1)}} \\
\vdots &  & \vdots & & \vdots &  & \vdots \\
M_{\varphiup(r+1),1;1,1} & \cdots & M_{\varphiup(r+1),1;1,N_1} & \cdots & M_{\varphiup(r+1),1;\varphiup(r+1),1} & \cdots & M_{\varphiup(r+1),1;\varphiup(r+1),N_{\varphiup(r+1)}} \\
\vdots &  & \vdots & & \vdots &  & \vdots \\
M_{\varphiup(r+1),N_{\varphiup(r+1)};1,1} & \cdots & M_{\varphiup(r+1),N_{\varphiup(r+1)};1,N_1} & \cdots & M_{\varphiup(r+1),N_{\varphiup(r+1)};\varphiup(r+1),1} & \cdots & M_{\varphiup(r+1),N_{\varphiup(r+1)};\varphiup(r+1),N_{\varphiup(r+1)}} \\
\end{pmatrix},
\end{align*}
whose blocks are defined as
\begin{align*}
\bM_{j,j';i,i'}=\bS(-\oa_{r+1}^{(j)}k_{j,j'})\bG(k_{i,i'},-\oa_{r+1}^{(j)}k_{j,j'})\bR(k_{i,i'}),
\end{align*}
in which
\begin{align*}
\bS(-\oa_{r+1}^{(j)}k_{j,j'})=
\begin{pmatrix}
\sigma(-\oa_{r+1}^{(j)}k_{j,j'})& & \\
\vdots& \ddots& \\
\frac{\partial_{k_{j,j'}}^{s_{j,j'}-1}\sigma(-\oa_{r+1}^{(j)}k_{j,j'})}{(s_{j,j'}-1)!}&\cdots &\sigma(-\oa_{r+1}^{(j)}k_{j,j'})
\end{pmatrix}, \quad
\bR(k_{i,i'})=
\begin{pmatrix}
\frac{\partial_{k_{i,i'}}^{s_{i,i'}-1}\rho(k_{i,i'})}{(s_{i,i'}-1)!}& \cdots& \rho(k_{i,i'})\\
\vdots& \iddots & \\
\rho(k_{i,i'})& &
\end{pmatrix},
\end{align*}
and
\begin{align*}
\bG(k_{i,i'},-\oa_{r+1}^{(j)}k_{j,j'})=(s_{j,j'}-1)!
\begin{pmatrix}
\frac{C_0^0}{k_{i,i'}-\oa_{r+1}^{(j)}k_{j,j'}}& \cdots &
\frac{C_{s_{i,i'}-1}^{0}}{(k_{i,i'}-\oa_{r+1}^{(j)}k_{j,j'})^{s_{i,i'}}{(-1)}^{s_{i,i'}-1}} \\
\vdots& & \vdots \\
\frac{{(-\oa_{r+1}^{(j)})}^{s_{j,j'}-1}C_{s_{j,j'}-1}^{s_{j,j'}-1}}{(k_{i,i'}-\oa_{r+1}^{(j)}k_{j,j'})^{s_{j,j'}}{(-1)}^{s_{j,j'}-1}} & \cdots & \frac{{(-\oa_{r+1}^{(j)})}^{s_{j,j'}-1}C_{s_{i,i'}+s_{j,j'}-2}^{s_{j,j'}-1}}{(k_{i,i'}-\oa_{r+1}^{(j)}k_{j,j'})^{s_{i,i'}+s_{j,j'}-1}{(-1)}^{s_{i,i'}+s_{j,j'}-2}}
\end{pmatrix}.
\end{align*}
With these symbols, we can reformulate equations \eqref{Ar1:U1} and \eqref{Ar1:u1} as
\begin{align*}
\bU=\tbuscr\bA\bD\bcscr'\bs\quad \hbox{and} \quad \tbuscr(I+\bA\bM)=\tbr\bcscr,
\end{align*}
respectively, which in turn implies
\begin{align*}
\bU=\tbr\bcscr(\bI+\bA\bM)^{-1}\bA\bD\bcscr'\bs;
\end{align*}
in other words, the $(\bar{i},\bar{j})$-entry takes the expression
\begin{align}\label{Ar1:Sol}
\bU^{(\bar{i},\bar{j})}=\tbr\bK^{(\bar{i})}(\bI+\bA\bM)^{-1}\bA\bD\bK'^{(\bar{j})}\bs,
\end{align}
where $\bK^{(\bar{i})}$ and $\bK'^{(\bar{j})}$ are diagonal block matrices as follows:
\begin{align*}
&\bK^{(\bar{i})}=\mathrm{diag}\left(\bK^{(\bar{i})}_{1,1},\cdots,\bK^{(\bar{i})}_{1,N_1};\cdots;
\bK^{(\bar{i})}_{\varphiup(r+1),1},\cdots,\bK^{(\bar{i})}_{\varphiup(r+1),N_{\varphiup(r+1)}}\right), \\
&\bK'^{(\bar{j})}=\mathrm{diag}\left(\bK'^{(\bar{j})}_{1,1},\cdots,\bK'^{(\bar{j})}_{1,N_1};\cdots;
\bK'^{(\bar{j})}_{\varphiup(r+1),1},\cdots,\bK'^{(\bar{j})}_{\varphiup(r+1),N_{\varphiup(r+1)}}\right),
\end{align*}
in which their respective blocks are defined by
\begin{align*}
\bK_{j,j'}^{(\bar{i})}=
\begin{pmatrix}
k_{j,j'}^{\bar{i}}  \\
\vdots& \ddots& \\
\frac{\partial_{k_{j,j'}}^{s_{j,j'}-1}k_{j,j'}^{\bar{i}}}{(s_{j,j'}-1)!}& \cdots&k_{j,j'}^{\bar{i}}
\end{pmatrix}
\quad \hbox{and} \quad
{\bK'}_{j,j'}^{(\bar{j})}=
\begin{pmatrix}
{(-\oa_{r+1}^{(j)}k_{j,j'})}^{\bar{j}} \\
\vdots & \ddots \\
\frac{\partial_{k_{j,j'}}^{s_{j,j'}-1}{(-\oa_{r+1}^{(j)}k_{j,j'})}^{\bar{j}}}{(s_{j,j'}-1)!}&\cdots &{(-\oa_{r+1}^{(j)}k_{j,j'})}^{\bar{j}}
\end{pmatrix},
\end{align*}
respectively.

\subsection{$A_{2r}^{(2)}$- and $D_{r+1}^{(2)}$-types}

The linear integral equations for the $A_{2r}^{(2)}$-type and $D_{r+1}^{(2)}$-type \ac{2DTL} can be written in a uniform way as \eqref{BN:Integral},
in which $\cN=2r+1$ and $\cN=2r+2$ for positive integers $r$ corresponding to the $A_{2r}^{(2)}$ and $D_{r+1}^{(2)}$ classes, respectively.
Similarly, the infinite matrix $\bU$ takes the form of \eqref{BN:Potential}.
To derive the Cauchy matrix solution, we still take the measure \eqref{Ar1:Pole} and replace this in \eqref{BN:Integral} and \eqref{BN:Potential} simultaneously.
By following a similar derivation, we end up with the general formula for $\bU^{(\bar{i},\bar{j})}$ as follows:
\begin{align}\label{BN:Sol}
\bU^{(\bar{i},\bar{j})}=(\tbr,\tbr')
\begin{pmatrix}
\bK^{(\bar{i})}& 0\\
0 & {\bK''}^{(\bar{i})}
\end{pmatrix}
\left[\bI+
\begin{pmatrix}
\bA& 0 \\
0& -\bA'
\end{pmatrix}
\begin{pmatrix}
{\bM}^{(1,1)}& {\bM}^{(1,2)} \\
{\bM}^{(2,1)}& {\bM}^{(2,2)}
\end{pmatrix}
\right]^{-1}
\begin{pmatrix}
\bA& 0 \\
0& -\bA'
\end{pmatrix}
\begin{pmatrix}
\bD& 0 \\
0& \bD
\end{pmatrix}	
\begin{pmatrix}
{\bK}'^{(\bar{j})}& 0\\
0 & {\bK'''}^{(\bar{j})}
\end{pmatrix}
\left(
\begin{array}{c}
\bs \\
\bs'
\end{array}
\right).
\end{align}
In formula \eqref{BN:Sol}, the $\bK$s are the diagonal block matrices given as follows:
\begin{align*}
&\bK^{(\bar{i})}=\mathrm{diag}\left(\bK^{(\bar{i})}_{1,1},\cdots,\bK^{(\bar{i})}_{1,N_1};\cdots;\bK^{(\bar{i})}_{\varphiup(\cN),1},\cdots,\bK^{(\bar{i})}_{\varphiup(\cN),N_{\varphiup(\cN)}}\right), \\
&{\bK'}^{(\bar{i})}=\mathrm{diag}\left({\bK'}^{(\bar{i})}_{1,1},\cdots,{\bK'}^{(\bar{i})}_{1,N_1};\cdots;{\bK'}^{(\bar{i})}_{\varphiup(\cN),1},\cdots,{\bK'}^{(\bar{i})}_{\varphiup(\cN),N_{\varphiup(\cN)}}\right), \\
&\bK''^{(\bar{i})}=\mathrm{diag}\left(\bK''^{(\bar{i})}_{1,1},\cdots,\bK''^{(\bar{i})}_{1,N_1};\cdots;\bK''^{(\bar{i})}_{\varphiup(\cN),1},\cdots,\bK''^{(\bar{i})}_{\varphiup(\cN),N_{\varphiup(\cN)}}\right), \\
&{\bK'''}^{(\bar{i})}=\mathrm{diag}\left({\bK'''}^{(\bar{i})}_{1,1},\cdots,{\bK'''}^{(\bar{i})}_{1,N_1};\cdots;{\bK'''}^{(\bar{i})}_{\varphiup(\cN),1},\cdots,{\bK'''}^{(\bar{i})}_{\varphiup(\cN),N_{\varphiup(\cN)}}\right),
\end{align*}
in which the blocks lower triangle matrices are defined as
\bse\label{BN:K}
\begin{align*}
\bK_{j,j'}^{(\bar{i})}=
\begin{pmatrix}
k_{j,j'}^{\bar{i}}& & \\
\vdots& \ddots& \\
\frac{\partial_{k_{j,j'}}^{s_{j,j'}-1}k_{j,j'}^{\bar{i}}}{(s_{j,j'}-1)!}& \cdots&k_{j,j'}^{\bar{i}}
\end{pmatrix},\quad
{\bK'}_{j,j'}^{(\bar{i})}=
\begin{pmatrix}
k_{j,j'}(-\oa_{\cN}^{(j)} k_{j,j'})^{\bar{i}}& & \\
\vdots& \ddots& \\
\frac{\partial_{k_{j,j'}}^{(s_{j,j'}-1)}k_{j,j'}(-\oa_{\cN}^{(j)} k_{j,j'})^{\bar{i}}}{(s_{j,j'}-1)!}& \cdots&k_{j,j'}(-\oa_{\cN}^{(j)} k_{j,j'})^{\bar{i}}
\end{pmatrix},
\end{align*}
and
\begin{align*}
{\bK''}_{j,j'}^{(\bar{i})}=
\begin{pmatrix}
{(-\oa_{\cN}^{(j)}k_{j,j'})}^{\bar{i}}& & \\
\vdots& \ddots& \\
\frac{\partial_{-\oa_{\cN}^{(j)}k_{j,j'}}^{s_{j,j'}-1}{(-\oa_{\cN}^{(j)}k_{j,j'})}^{\bar{i}}}{(s_{j,j'}-1)!}&\cdots &{(-\oa_{\cN}^{(j)}k_{j,j'})}^{\bar{i}}
\end{pmatrix}, \quad
{\bK'''}_{j,j'}^{(\bar{i})}=
\begin{pmatrix}
{-\oa_{\cN}^{(j)}k^{\bar{i}+1}_{j,j'}}& & \\
\vdots& \ddots& \\
\frac{\partial_{k_{j,j'}}^{(s_{j,j'}-1)}(-\oa_{\cN}^{(j)}{k^{\bar{i}+1}_{j,j'}})}{(s_{j,j'}-1)!}&\cdots &{-\oa_{\cN}^{(j)}k^{\bar{i}+1}_{j,j'}}
\end{pmatrix},
\end{align*}
respectively.
\ese
The $\bA$, $\bA'$ take their respective forms of
\begin{align}\label{BN:A}
\bA=\mathrm{diag}\left(\bA_{1,1},\cdots,\bA_{1,N_1};\cdots;\bA_{\varphiup(\cN),1},\cdots,\bA_{\varphiup(\cN),N_{\varphiup(\cN)}}\right) \quad \hbox{and} \quad
\bA'=\mathrm{diag}\left(\bA'_{1,1},\cdots,\bA'_{1,N_1};\cdots;\bA'_{\varphiup(\cN),1},\cdots,\bA'_{\varphiup(\cN),N_{\varphiup(\cN)}}\right),
\end{align}
where
\begin{align*}
\bA_{j,j'}=A_{j,j'}\bI_{s_{j,j'}\times s_{j,j'}} \quad \hbox{and} \quad
\bA'_{j,j'}=\mathrm{diag}\left(A_{j,j'}{(-\oa_{\cN})}^{s_{j,j'}-1},\cdots,A_{j,j'}\right).
\end{align*}
The $\bD$ is a diagonal matrix in the form of
\begin{align}\label{bN:D}
\bD=\left(\bD_{1,1},\cdots,\bD_{1,N_1};\cdots;\bD_{\varphiup(r+1),1},\cdots,\bD_{\varphiup(r+1),N_{\varphiup(r+1)}}\right), \quad \hbox{with} \quad
\bD_{j,j'}=(s_{j,j'}-1)!\bI_{s_{j,j'}\times s_{j,j'}}.
\end{align}
The $\tbr$ and $\tbr'$ are finite row vectors defined as
\begin{align}\label{BN:r}
\tbr=\left(\tbr_{1,1},\cdots,\tbr_{1,N_1};\cdots;\tbr_{{\varphiup(\cN)},1},\cdots,\tbr_{\varphiup(\cN),N_{\varphiup(\cN)}}\right) \quad \hbox{and} \quad
\tbr'=\left(\tbr'_{1,1},\cdots,\tbr'_{1,N_1};\cdots;\tbr'_{{\varphiup(\cN)},1},\cdots,{\tbr'}_{\varphiup(\cN),N_{\varphiup(\cN)}}\right),
\end{align}
respectively, in which
\begin{align*}
\tbr_{j,j'}=\left(\frac{\partial_{k_{j,j'}}^{s_{j,j'}-1}\rho(k_{j,j'})}{(s_{j,j'}-1)!},\cdots,\rho(k_{j,j'})\right), \quad
\tbr'_{j,j'}=\left(\frac{\partial_{-\oa_{\cN}^{(j)}k_{j,j'}}^{s_{j,j'}-1}\rho(-\oa_{\cN}^{(j)}k_{j,j'})}{(s_{j,j'}-1)!},\cdots,\rho(-\oa_{\cN}^{(j)}k_{j,j'})\right).
\end{align*}
The $\bs$ and $\bs'$ are finite column vectors given by
\begin{align}\label{BN:s}
\bs={}^{t\!}\left(\tbs_{1,1},\cdots,\tbs_{1,N_1};\cdots;\tbs_{{\varphiup(\cN)},1},\cdots,\tbs_{\varphiup(\cN),N_{\varphiup(\cN)}}\right) \quad \hbox{and} \quad
\bs'={}^{t\!}\left(\tbs'_{1,1},\cdots,\tbs'_{1,N_1};\cdots;\tbs'_{{\varphiup(\cN)},1},\cdots,\tbs'_{\varphiup(\cN),N_{\varphiup(\cN)}}\right),
\end{align}
respectively, where
\begin{align*}
\tbs_{j,j'}=\left(\sigma(-\oa_{\cN}^{(j)}k_{j,j'}),\cdots,\frac{\partial_{k_{j,j'}}^{s_{j,j'}-1}\sigma(-\oa_{\cN}^{(j)}k_{j,j'})}{(s_{j,j'}-1)!}\right), \quad
\tbs'_{j,j'}=\left(\sigma(k_{j,j'}),\cdots,\frac{\partial_{k_{j,j'}}^{s_{j,j'}-1}\sigma{(k_{j,j'})}}{(s_{j,j'}-1)!}\right).
\end{align*}
The $\bM$ is a block Cauchy matrix with its blocks $\bM^{(\alpha,\beta)}$ defined by
\begin{align}\label{BN:M}
\bM^{(\alpha,\beta)}=
\begin{pmatrix}
M_{1,1;1,1}^{(\alpha,\beta)} & \cdots & M_{1,1;1,N_1}^{(\alpha,\beta)} & \cdots & M_{1,1;\varphiup(\cN),1}^{(\alpha,\beta)} & \cdots & M_{1,1;\varphiup(\cN),N_{\varphiup(\cN)}}^{(\alpha,\beta)} \\
\vdots &  & \vdots & & \vdots &  & \vdots  \\
M_{1,N_1;1,1}^{(\alpha,\beta)} & \cdots & M_{1,N_1;1,N_1}^{(\alpha,\beta)} & \cdots & M_{1,N_1;\varphiup(\cN),1}^{(\alpha,\beta)} & \cdots & M_{1,N_1;\varphiup(\cN),N_{\varphiup(\cN)}}^{(\alpha,\beta)} \\
\vdots &  & \vdots & & \vdots &  & \vdots \\
M_{\varphiup(\cN),1;1,1}^{(\alpha,\beta)} & \cdots & M_{\varphiup(\cN),1;1,N_1}^{(\alpha,\beta)} & \cdots & M_{\varphiup(\cN),1;\varphiup(\cN),1}^{(\alpha,\beta)} & \cdots & M_{\varphiup(\cN),1;\varphiup(\cN),N_{\varphiup(\cN)}}^{(\alpha,\beta)} \\
\vdots &  & \vdots & & \vdots &  & \vdots \\
M_{\varphiup(\cN),N_{\varphiup(\cN)};1,1}^{(\alpha,\beta)} & \cdots & M_{\varphiup(\cN),N_{\varphiup(\cN)};1,N_1}^{(\alpha,\beta)} & \cdots & M_{\varphiup(\cN),N_{\varphiup(\cN)};\varphiup(\cN),1}^{(\alpha,\beta)} & \cdots & M_{\varphiup(\cN),N_{\varphiup(\cN)};\varphiup(\cN),N_{\varphiup(\cN)}}^{(\alpha,\beta)} \\
\end{pmatrix},
\end{align}
for $\alpha,\beta\in\{1,2\}$. Here the ${\bM}^{(1,1)}_{j,j';i,i'}$ takes the form of
\begin{align}\label{BN:M11}
{\bM}^{(1,1)}_{j,j';i,i'}=\bS{(-\oa_{\cN}^{(j)} k_{j,j'})}{\bG}{(k_{i,i'},-\oa_{\cN}^{(j)}k_{j,j'})}\bR{(k_{i,i'})},
\end{align}
in which
\begin{align}\label{BN:RS11}
\bS{(-\oa_{\cN}^{(j)} k_{j,j'})}=
\begin{pmatrix}
\sigma(-\oa_{\cN}^{(j)}k_{j,j'})& & \\
\vdots& \ddots& \\
\frac{\partial_{k_{j,j'}}^{(s_{j,j'}-1)}\sigma(-\oa_{\cN}^{(j)}k_{j,j'})}{(s_{j,j'}-1)!}&\cdots &\sigma(-\oa_{\cN}^{(j)}k_{j,j'})
\end{pmatrix},\quad
\bR{(k_{i,i'})}=
\begin{pmatrix}
\frac{\partial_{k_{i,i'}}^{s_{i,i'}-1}\rho(k_{i,i'})}{(s_{i,i'}-1)!}& \cdots& \rho(k_{i,i'})\\
\vdots& \iddots & \\
\rho(k_{i,i'})& &
\end{pmatrix},
\end{align}
and
\begin{align*}
{\bG}{(k_{i,i'},-\oa_{\cN}^{(j)}k_{j,j'})}=(s_{j,j'}-1)!
\begin{pmatrix}
\frac{k_{j,j'}C_0^0}{k_{i,i'}-\oa_{\cN}^{(j)}k_{j,j'}}& \cdots &
\frac{k_{j,j'}C_{s_{i,i'}-1}^0}{(k_{i,i'}-\oa_{\cN}^{(j)}k_{j,j'})^{s_{i,i'}}{(-1)}^{s_{i,i'}-1}} \\
\vdots& & \vdots \\
\frac{k_{j,j'}{(-\oa_{\cN}^{(j)})}^{s_{j,j'}-1}C_{s_{j,j'}-1}^{s_{j,j'}-1}}{(k_{i,i'}-\oa_{\cN}^{(j)}k_{j,j'})^{s_{j,j'}}{(-1)}^{s_{j,j'}-1}}
+\frac{{(-\oa_{\cN}^{(j)})}^{s_{j,j'}-2}C_{s_{j,j'}-2}^{s_{j,j'}-2}}{(k_{i,i'}-\oa_{\cN}^{(j)}k_{j,j'})^{s_{j,j'}-1}{(-1)}^{s_{j,j'}-2}} & \cdots & \spadesuit
\end{pmatrix}
\end{align*}
for
\begin{align*}
\spadesuit=\frac{k_{j,j'}{(-\oa_{\cN}^{(j)})}^{s_{j,j'}-1}C_{s_{i,i'}+s_{j,j'}-2}^{s_{j,j'}-1}}{(k_{i,i'}-\oa_{\cN}^{(j)}k_{j,j'})^{s_{i,i'}+s_{j,j'}-1}{(-1)}^{s_{i,i'}+s_{j,j'}-2}}
+\frac{{(-\oa_{\cN}^{(j)})}^{s_{j,j'}-2}C_{s_{i,i'}+s_{j,j'}-3}^{s_{j,j'}-2}}{(k_{i,i'}-\oa_{\cN}^{(j)}k_{j,j'})^{s_{i,i'}+s_{j,j'}-2}{(-1)}^{s_{i,i'}+s_{j,j'}-3}};
\end{align*}
the ${\bM}^{(1,2)}_{j,j';i,i'}$ is defined as
\begin{align}\label{BN:M12}
{\bM}^{(1,2)}_{j,j';i,i'}=\bS{(-\oa_{\cN}^{(j)} k_{j,j'})}{\bG}{(-\oa_{\cN}^{(j)} k_{i,i'},-\oa_{\cN}^{(j)} k_{j,j'})}\bR(-\oa_{\cN}^{(j)} k_{i,i'})
\end{align}
in which
\begin{align}\label{BN:RS12}
\bS{(-\oa_{\cN}^{(j)} k_{j,j'})}=
\begin{pmatrix}
\sigma(-\oa_{\cN}^{(j)}k_{j,j'})& & \\
\vdots& \ddots& \\
\frac{\partial_{k_{j,j'}}^{(s_{j,j'}-1)}\sigma(-\oa_{\cN}^{(j)}k_{j,j'})}{(s_{j,j'}-1)!}&\cdots &\sigma(-\oa_{\cN}^{(j)}k_{j,j'})
\end{pmatrix},\quad
\bR{(-\oa_{\cN}^{(j)} k_{i,i'})}=
\begin{pmatrix}
\frac{\partial_{-\oa_{\cN}^{(j)}k_{i,i'}}^{s_{i,i'}-1}\rho(-\oa_{\cN}^{(j)}k_{i,i'})}{(s_{i,i'}-1)!}& \cdots& \rho( -\oa_{\cN}^{(j)}k_{i,i'})\\
\vdots& \iddots & \\
\rho(-\oa_{\cN}^{(j)}k_{i,i'})& &
\end{pmatrix},
\end{align}
and
\begin{align*}
{\bG}{(-\oa_{\cN}^{(j)} k_{i,i'},-\oa_{\cN}^{(j)} k_{j,j'})}=(s_{j,j'}-1)!
\begin{pmatrix}
\frac{k_{j,j'}C_0^0}{-\oa_{\cN}^{(j)}k_{i,i'}-\oa_{\cN}^{(j)}k_{j,j'}}& \cdots &
\frac{k_{j,j'}C_{s_{i,i'}-1}^0}{(-\oa_{\cN}^{(j)}k_{i,i'}-\oa_{\cN}^{(j)}k_{j,j'})^{s_{i,i'}}{(-1)}^{s_{i,i'}-1}} \\
\vdots& & \vdots \\
\frac{k_{j,j'}{(-\oa_{\cN}^{(j)})}^{s_{j,j'}-1}C_{s_{j,j'}-1}^{s_{j,j'}-1}}{(-\oa_{\cN}^{(j)}k_{i,i'}-\oa_{\cN}^{(j)}k_{j,j'})^{s_{j,j'}}{(-1)}^{s_{j,j'}-1}}
+\frac{{(-\oa_{\cN}^{(j)})}^{s_{j,j'}-2}C_{s_{j,j'}-2}^{s_{j,j'}-2}}{(-\oa_{\cN}^{(j)}k_{i,i'}-\oa_{\cN}^{(j)}k_{j,j'})^{s_{j,j'}-1}{(-1)}^{s_{j,j'}-2}} & \cdots & \heartsuit
\end{pmatrix}
\end{align*}
for
\begin{align*}
\heartsuit=\frac{k_{j,j'}{(-\oa_{\cN}^{(j)})}^{s_{j,j'}-1}C_{s_{i,i'}+s_{j,j'}-2}^{s_{j,j'}-1}}{(-\oa_{\cN}^{(j)}k_{i,i'}-\oa_{\cN}^{(j)}k_{j,j'})^{s_{i,i'}+s_{j,j'}-1}{(-1)}^{s_{i,i'}+s_{j,j'}-2}}
+\frac{{(-\oa_{\cN}^{(j)})}^{s_{j,j'}-2}C_{s_{i,i'}+s_{j,j'}-3}^{s_{j,j'}-2}}{(-\oa_{\cN}^{(j)}k_{i,i'}-\oa_{\cN}^{(j)}k_{j,j'})^{s_{i,i'}+s_{j,j'}-2}{(-1)}^{s_{i,i'}+s_{j,j'}-3}};
\end{align*}
the ${\bM}^{(2,1)}_{j,j';i,i'}$ is expressed by
\begin{align}\label{BN:M21}
{\bM}^{(2,1)}_{j,j';i,i'}=\bS{(k_{j,j'})}{\bG}{ (k_{i,i'},k_{j,j'})}\bR{(k_{i,i'})},
\end{align}
where
\begin{align}\label{BN:RS21}
\bS{(k_{j,j'})}=
\begin{pmatrix}
\sigma(k_{j,j'})& & \\
\vdots& \ddots& \\
\frac{\partial_{k_{j,j'}}^{(s_{j,j'}-1)}\sigma(k_{j,j'})}{(s_{j,j'}-1)!}&\cdots &\sigma(k_{j,j'})
\end{pmatrix},\quad
\bR{(k_{i,i'})}=
\begin{pmatrix}
\frac{\partial_{k_{i,i'}}^{s_{i,i'}-1}\rho(k_{i,i'})}{(s_{i,i'}-1)!}& \cdots& \rho(k_{i,i'})\\
\vdots& \iddots & \\
\rho(k_{i,i'})& &
\end{pmatrix},
\end{align}
and
\begin{align*}
{\bG}{ (k_{i,i'},k_{j,j'})}=(s_{j,j'}-1)!
\begin{pmatrix}
\frac{-\oa_{\cN}^{(j)}k_{j,j'}C_0^0}{k_{i,i'}+k_{j,j'}}& \cdots &
\frac{-\oa_{\cN}^{(j)}k_{j,j'}C_{s_{i,i'}-1}^0}{(k_{i,i'}+k_{j,j'})^{s_{i,i'}}{(-1)}^{s_{i,i'}-1}} \\
\vdots& & \vdots \\
\frac{-\oa_{\cN}^{(j)}k_{j,j'}C_{s_{j,j'}-1}^{s_{j,j'}-1}}{(k_{i,i'}+k_{j,j'})^{s_{j,j'}}{(-1)}^{s_{j,j'}-1}}
+\frac{\oa_{\cN}^{(j)}C_{s_{j,j'}-2}^{s_{j,j'}-2}}{(k_{i,i'}+k_{j,j'})^{s_{j,j'}-1}{(-1)}^{s_{j,j'}-1}} & \cdots & \clubsuit
\end{pmatrix}
\end{align*}
for
\begin{align*}
\clubsuit=\frac{-\oa_{\cN}^{(j)}k_{j,j'}C_{s_{i,i'}+s_{j,j'}-2}^{s_{j,j'}-1}}{(k_{i,i'}+k_{j,j'})^{s_{i,i'}+s_{j,j'}-1}{(-1)}^{s_{i,i'}+s_{j,j'}-2}}
+\frac{\oa_{\cN}^{(j)}C_{s_{i,i'}+s_{j,j'}-3}^{s_{j,j'}-2}}{(k_{i,i'}+k_{j,j'})^{s_{i,i'}+s_{j,j'}-2}{(-1)}^{s_{i,i'}+s_{j,j'}-3}};
\end{align*}
and finally ${\bM}^{(2,2)}_{j,j';i,i'}$ is given as
\begin{align}\label{BN:M22}
{\bM}^{(2,2)}_{j,j';i,i'}=\bS{(k_{j,j'})}{\bG}{(-\oa_{\cN}^{(j)} k_{i,i'},k_{j,j'})}\bR{(-\oa_{\cN}^{(j)} k_{i,i'})},
\end{align}
where
\begin{align}\label{BN:RS22}
\bS{(k_{j,j'})}=
\begin{pmatrix}
\sigma(k_{j,j'})& & \\
\vdots& \ddots& \\
\frac{\partial_{k_{j,j'}}^{(s_{j,j'}-1)}\sigma(k_{j,j'})}{(s_{j,j'}-1)!}&\cdots &\sigma(k_{j,j'})
\end{pmatrix},\quad
\bR{(-\oa_{\cN}^{(j)} k_{i,i'})}=
\begin{pmatrix}
\frac{\partial_{-\oa_{\cN}^{(j)}k_{i,i'}}^{s_{i,i'}-1}\rho(-\oa_{\cN}^{(j)}k_{i,i'})}{(s_{i,i'}-1)!}& \cdots& \rho(-\oa_{\cN}^{(j)}k_{i,i'})\\
\vdots& \iddots & \\
\rho(-\oa_{\cN}^{(j)}k_{i,i'})& &
\end{pmatrix},
\end{align}
and
\begin{align*}
&{\bG}(-\oa_{\cN}^{(j)} k_{i,i'},k_{j,j'})=(s_{j,j'}-1)!
\begin{pmatrix}
\frac{-\oa_{\cN}^{(j)}k_{j,j'}C_0^0}{-\oa_{\cN}^{(j)}k_{i,i'}+k_{j,j'}}& \cdots &
\frac{-\oa_{\cN}^{(j)}k_{j,j'}C_{s_{i,i'}-1}^0}{(-\oa_{\cN}^{(j)}k_{i,i'}+k_{j,j'})^{s_{i,i'}}{(-1)}^{s_{i,i'}-1}} \\
\vdots& & \vdots \\
\frac{-\oa_{\cN}^{(j)}k_{j,j'}C_{s_{j,j'}-1}^{s_{j,j'}-1}}{(-\oa_{\cN}^{(j)}k_{i,i'}+k_{j,j'})^{s_{j,j'}}{(-1)}^{s_{j,j'}-1}}
+\frac{\oa_{\cN}^{(j)}C_{s_{j,j'}-2}^{s_{j,j'}-2}}{(-\oa_{\cN}^{(j)}k_{i,i'}+k_{j,j'})^{s_{j,j'}-1}{(-1)}^{s_{j,j'}-1}} & \cdots & \diamondsuit
\end{pmatrix}
\end{align*}
for
\begin{align*}
\diamondsuit =\frac{-\oa_{\cN}^{(j)}k_{j,j'}C_{s_{i,i'}+s_{j,j'}-2}^{s_{j,j'}-1}}{(-\oa_{\cN}^{(j)}k_{i,i'}+k_{j,j'})^{s_{i,i'}+s_{j,j'}-1}{(-1)}^{s_{i,i'}+s_{j,j'}-2}}
+\frac{\oa_{\cN}^{(j)}C_{s_{i,i'}+s_{j,j'}-3}^{s_{j,j'}-2}}{(-\oa_{\cN}^{(j)}k_{i,i'}+k_{j,j'})^{s_{i,i'}+s_{j,j'}-2}{(-1)}^{s_{i,i'}+s_{j,j'}-3}}.
\end{align*}

\subsection{$A_{2r}^{(2)}$- and $C_{r}^{(1)}$-types}

The linear integral equations and the potentials in the $A_{2r}^{(2)}$ and $C_r^{(1)}$ classes have the unified forms \eqref{CN:Integral} and
\eqref{CN:Potential}, respectively.
Taking the same measure as \eqref{A:Pole} and performing this on \eqref{CN:Integral} and \eqref{CN:Potential},
we are able to construct the Cauchy matrix formula of $\bU^{(\bar{i},\bar{j})}$, by following a similar derivation to that in the previous cases,
which takes the form of
\begin{align}\label{CN:Sol}
\bU^{(\bar{i},\bar{j})}=(\tbr,\tbr')
\begin{pmatrix}
\bK^{(\bar{i})}& 0\\
0 & \bK''^{(\bar{i})}
\end{pmatrix}
\left[\bI+
\begin{pmatrix}
\bA& 0 \\
0& \bA'
\end{pmatrix}
\begin{pmatrix}
\bM^{(1,1)}& \bM^{(1,2)} \\
\bM^{(2,1)}& \bM^{(2,2)}
\end{pmatrix}
\right]^{-1}
\begin{pmatrix}
\bA& 0 \\
0& \bA'
\end{pmatrix}
\begin{pmatrix}
\bD& 0 \\
0& \bD
\end{pmatrix}
\begin{pmatrix}
\bK'^{(\bar{j})}& 0\\
0 & \bK^{(\bar{j})}
\end{pmatrix}
\left(
\begin{array}{c}
\bs \\
\bs'
\end{array}
\right).
\end{align}
Here $\bA$, $\bA'$, $\bD$, $\br$, $\br'$, $\bs$ as well as $\bs'$ are exactly the same as those given in \eqref{BN:A}, \eqref{bN:D}, \eqref{BN:r} and \eqref{BN:s}.
The $\bK^{(\bar{i})}$, $\bK'^{(\bar{i})}$ and $\bK''^{(\bar{i})}$ are given as follows:
\begin{align*}
&\bK^{(\bar{i})}=\mathrm{diag}(\bK^{(\bar{i})}_{1,1},\cdots,\bK^{(\bar{i})}_{1,N_1};\cdots;\bK^{(\bar{i})}_{\varphiup(\cN),1},\cdots,\bK^{(\bar{i})}_{\varphiup(\cN),N_{\varphiup(\cN)}}), \\
&\bK'^{(\bar{i})}=\mathrm{diag}(\bK'^{(\bar{i})}_{1,1},\cdots,\bK'^{(\bar{i})}_{1,N_1};\cdots;\bK'^{(\bar{i})}_{\varphiup(\cN),1},\cdots,\bK'^{(\bar{i})}_{\varphiup(\cN),N_{\varphiup(\cN)}}), \\
&\bK''^{(\bar{i})}=\mathrm{diag}(\bK''^{(\bar{i})}_{1,1},\cdots,\bK''^{(\bar{i})}_{1,N_1};\cdots;\bK''^{(\bar{i})}_{\varphiup(\cN),1},\cdots,\bK''^{(\bar{i})}_{\varphiup(\cN),N_{\varphiup(\cN)}}),
\end{align*}
where
\begin{align*}
\bK_{j,j'}^{(\bar{i})}=
\begin{pmatrix}
k_{j,j'}^{\bar{i}}& & \\
\vdots& \ddots& \\
\frac{\partial_{k_{j,j'}}^{s_{j,j'}-1}k_{j,j'}^{\bar{i}}}{(s_{j,j'}-1)!}& \cdots&k_{j,j'}^{\bar{i}}
\end{pmatrix},\quad
{\bK'}_{j,j'}^{(\bar{i})}=
\begin{pmatrix}
{(-\oa_{\cN}^{(j)}k_{j,j'})}^{\bar{i}}& & \\
\vdots& \ddots& \\
\frac{\partial_{k_{j,j'}}^{s_{j,j'}-1}{(-\oa_{\cN}^{(j)}k_{j,j'})}^{\bar{i}}}{(s_{j,j'}-1)!}&\cdots &{(-\oa_{\cN}^{(j)}k_{j,j'})}^{\bar{i}}
\end{pmatrix},
\end{align*}
and
\begin{align*}
{\bK''}_{j,j'}^{(\bar{i})}=
\begin{pmatrix}
{(-\oa_{\cN}^{(j)}k_{j,j'})}^{\bar{i}}& & \\
\vdots& \ddots& \\
\frac{\partial_{-\oa_{\cN}^{(j)}k_{j,j'}}^{s_{j,j'}-1}{(-\oa_{\cN}^{(j)}k_{j,j'})}^{\bar{i}}}{(s_{j,j'}-1)!}&\cdots &{(-\oa_{\cN}^{(j)}k_{j,j'})}^{\bar{i}}
\end{pmatrix}.
\end{align*}
The Cauchy matrices $\bM^{(\alpha,\beta)}$ take their respect forms of \eqref{BN:M},
in which $\bM^{(\alpha,\beta)}_{j,j';i,i'}$ are defined in \eqref{BN:M11}, \eqref{BN:M12}, \eqref{BN:M21} and \eqref{BN:M22}, respectively,
with the same $\bR$ and $\bS$ matrices listed in \eqref{BN:RS11}, \eqref{BN:RS12}, \eqref{BN:RS21} and \eqref{BN:RS22},
but different $\bG$ matrices as follows:
\begin{align*}
\bG{(k_{i,i'},-\oa_{\cN}^{(j)} k_{j,j'})}&=(s_{j,j'}-1)!
\begin{pmatrix}
\frac{C_0^0}{k_{i,i'}-\oa_{\cN}^{(j)}k_{j,j'}}& \cdots &
\frac{C_{s_{i,i'}-1}^{0}}{(k_{i,i'}-\oa_{\cN}^{(j)}k_{j,j'})^{s_{i,i'}}{(-1)}^{s_{i,i'}-1}} \\
\vdots& & \vdots \\
\frac{{(-\oa_{\cN}^{(j)})}^{s_{j,j'}-1}C_{s_{j,j'}-1}^{s_{j,j'}-1}}{(k_{i,i'}-\oa_{\cN}^{(j)}k_{j,j'})^{s_{j,j'}}{(-1)}^{s_{j,j'}-1}} & \cdots & \frac{{(-\oa_{\cN}^{(j)})}^{s_{j,j'}-1}C_{s_{i,i'}+s_{j,j'}-2}^{s_{j,j'}-1}}{(k_{i,i'}-\oa_{\cN}^{(j)}k_{j,j'})^{s_{i,i'}+s_{j,j'}-1}{(-1)}^{s_{i,i'}+s_{j,j'}-2}}
\end{pmatrix}, \\
\bG{(-\oa_{\cN}^{(j)} k_{i,i'},-\oa_{\cN}^{(j)} k_{j,j'})}&=(s_{j,j'}-1)!
\begin{pmatrix}
\frac{C_0^0}{-\oa_{\cN}^{(j)}k_{i,i'}-\oa_{\cN}^{(j)}k_{j,j'}}& \cdots &
\frac{C_{s_{i,i'}-1}^{0}}{(-\oa_{\cN}^{(j)}k_{i,i'}-\oa_{\cN}^{(j)}k_{j,j'})^{s_{i,i'}}{(-1)}^{s_{i,i'}-1}} \\
\vdots& & \vdots \\
\frac{{(-\oa_{\cN}^{(j)})}^{s_{j,j'}-1}C_{s_{j,j'}-1}^{s_{j,j'}-1}}{(-\oa_{\cN}^{(j)}k_{i,i'}-\oa_{\cN}^{(j)}k_{j,j'})^{s_{j,j'}}{(-1)}^{s_{j,j'}-1}} & \cdots & \frac{{(-\oa_{\cN}^{(j)})}^{s_{j,j'}-1}C_{s_{i,i'}+s_{j,j'}-2}^{s_{j,j'}-1}}{(-\oa_{\cN}^{(j)}k_{i,i'}-\oa_{\cN}^{(j)}k_{j,j'})^{s_{i,i'}+s_{j,j'}-1}{(-1)}^{s_{i,i'}+s_{j,j'}-2}}
\end{pmatrix},\\
\bG'{(k_{i,i'},k_{j,j'})}&=(s_{j,j'}-1)!
\begin{pmatrix}
\frac{C_0^0}{k_{i,i'}+k_{j,j'}}& \cdots &
\frac{C_{s_{i,i'}-1}^0}{(k_{i,i'}+k_{j,j'})^{s_{i,i'}}{(-1)}^{s_{i,i'}-1}} \\
\vdots& & \vdots \\
\frac{C_{s_{j,j'}-1}^{s_{j,j'}-1}}{(k_{i,i'}+k_{j,j'})^{s_{j,j'}}{(-1)}^{s_{j,j'}-1}} & \cdots & \frac{C_{s_{i,i'}+s_{j,j'}-2}^{s_{j,j'}-1}}{(k_{i,i'}+k_{j,j'})^{s_{i,i'}+s_{j,j'}-1}{(-1)}^{s_{i,i'}+s_{j,j'}-2}}
\end{pmatrix}, \\
\bG{(-\oa_{\cN}^{(j)} k_{i,i'},k_{j,j'})}&=(s_{j,j'}-1)!
\begin{pmatrix}
\frac{C_0^0}{-\oa_{\cN}^{(j)}k_{i,i'}+k_{j,j'}}& \cdots &
\frac{C_{s_{i,i'}-1}^0}{(-\oa_{\cN}^{(j)}k_{i,i'}+k_{j,j'})^{s_{i,i'}}{(-1)}^{s_{i,i'}-1}} \\
\vdots& & \vdots \\
\frac{C_{s_{j,j'}-1}^{s_{j,j'}-1}}{(-\oa_{\cN}^{(j)}k_{i,i'}+k_{j,j'})^{s_{j,j'}}{(-1)}^{s_{j,j'}-1}} & \cdots & \frac{C_{s_{i,i'}+s_{j,j'}-2}^{s_{j,j'}-1}}{(-\oa_{\cN}^{(j)}k_{i,i'}+k_{j,j'})^{s_{i,i'}+s_{j,j'}-1}{(-1)}^{s_{i,i'}+s_{j,j'}-2}}
\end{pmatrix}.
\end{align*}
Then the expression $\varphi=\ln\left(1-\bU^{(0,-1)}\right)$ provides the Cauchy matrix solutions to the corresponding \ac{2D} Toda-type equations.
The $\cN=2r$ and $\cN=2r+1$ cases for positive integers $r$ correspond to the \ac{2D} Toda systems of $C_{r}^{(1)}$- and $A_{2r}^{(2)}$-types, respectively.

\subsection{Examples}

We have presented all the general formulae for Cauchy matrix solutions to the \ac{2DTL}s.
Below we take the three scalar equations, namely the $A_\infty$-type \ac{2D} Toda equation, the sinh--Gordon equation and the Tzitzeica equation as examples,
to show how soliton solution, multi-pole solution and their mixture are obtained as special cases of the Cauchy matrix solutions.

\begin{example}\label{E:2DTL}
The first example is the \ac{2D} Toda equation of $A_\infty$-type. We consider a special case of \eqref{A:Pole} as follows:
\begin{align*}
\mathrm{d}\zeta(k,k')=\sum_{j=1}^{2}\sum_{j=1}^{3}A_{j,j'}\frac{1}{(2\pi \mathrm{i})^2}\frac{1}{k-k_j}\frac{1}{k'-k'_{j'}}\rd k\rd k',
\end{align*}
namely the double integration has only first-order singularities $k_j$ for $j=1,2$ and $k'_{j'}$ for $j'=1,2,3$ on the $k$- and $k'$- planes,
located in the contours $\Gamma$ and $\Gamma'$, respectively.
This results in the expression of $\bU_n^{(0,-1)}$ given by
\begin{align*}
\bU_n^{(0,-1)}=\tbr(\bI+\bA\bM)^{-1}\bA{\bK'^{(-1)}}\bs,
\end{align*}
where the matrices $\bA$, $\bM$ and $K'^{(-1)}$ and the vectors $\br$ and $\bs$ are defined as
\begin{align*}
\bA=
\begin{pmatrix}
A_{1,1} & A_{1,2} & A_{1,3} \\
A_{2,1} & A_{2,2} & A_{2,3}
\end{pmatrix},
\quad
\bM=
\begin{pmatrix}
\frac{\rho_n(k_1)\sigma_n(k'_{1})}{k_1+k'_{1}} & \frac{\rho_n(k_2)\sigma_n(k'_{1})}{k_2+k'_{1}} \\
\frac{\rho_n(k_1)\sigma_n(k'_{2})}{k_1+k'_{2}} & \frac{\rho_n(k_2)\sigma_n(k'_{2})}{k_2+k'_{2}} \\
\frac{\rho_n(k_1)\sigma_n(k'_{3})}{k_1+k'_{3}} & \frac{\rho_n(k_2)\sigma_n(k'_{3})}{k_2+k'_{3}}
\end{pmatrix},
\quad
\bK'^{(-1)}=\diag\left({k'}_1^{-1},{k'}_2^{-1},{k'}_3^{-1}\right),
\end{align*}
and
\begin{align*}
\tbr=\left(\rho_n(k_1),\rho_n(k_2)\right), \quad \hbox{as well as} \quad \bs={}^{t\!}\left(\sigma_n(k_1),\sigma_n(k_2),\sigma_n(k_3)\right),
\end{align*}
respectively, with $\rho_n(k)$ and $\sigma_n(k')$ defined by \eqref{PWF}.
We refer to the corresponding $\varphi_n$ defined in \eqref{2DTL:Sol} as the $(2,3)$-soliton solution for \eqref{A:NL}.
\end{example}
\begin{example}\label{E:sG}
In the $A_1^{(1)}$ class, since there is only one primitive square root of unit $\omega_2^{(1)}=-1$,
we consider a special measure
\begin{align*}
\mathrm{d}\lambda_1(k)=A_{1,1}\frac{1}{2\pi\mathrm{i}}\frac{1}{(k-k_{1,1})^2}\rd k,	
\end{align*}
namely, we let the contour $\Gamma_1$ on the $k$-plane contain only a double-pole $k_{1,1}$ inside.
Then the general formula \eqref{Ar1:Sol} provides
\begin{align*}
\bU_0^{(0,-1)}=\tbr(\bI+\bA\bM)^{-1}\bA{\bK'^{(-1)}}\bs,
\end{align*}
in which the matrices $\bA$, $\bM$ and $\bK^{(-1)}$ are given as follows:
\begin{align*}
\bA=
\begin{pmatrix}
A_{1,1} & 0 \\
0 & A_{1,1}
\end{pmatrix},
\quad
\bM=
\begin{pmatrix}
\sigma_0(k_{1,1}) & 0 \\
\partial_{k_{1,1}}\sigma_0(k_{1,1}) & \sigma_0(k_{1,1})
\end{pmatrix}
\begin{pmatrix}
\frac{1}{k_{1,1}+k_{1,1}} & -\frac{1}{(k_{1,1}+k_{1,1})^2} \\
-\frac{1}{(k_{1,1}+k_{1,1})^2} & \frac{2}{(k_{1,1}+k_{1,1})^3}
\end{pmatrix}
\begin{pmatrix}
\partial_{k_{1,1}}\rho_0(k_{1,1})& \rho_0(k_{1,1}) \\
\rho_0(k_{1,1}) & 0
\end{pmatrix},
\quad
\bK'^{(-1)}=
\begin{pmatrix}
k^{-1}_{1,1}& 0 \\
-k^{-2}_{1,1} & k^{-1}_{1,1}
\end{pmatrix},
\end{align*}
and the vectors $\br$ and $\bs$ are defined as
\begin{align*}
\tbr=\left(\partial_{k_{1,1}}\rho_0(k_{1,1}),\rho_0(k_{1,1})\right) \quad \hbox{and} \quad \bs={}^{t\!}\left(\sigma_0(k_{1,1}),\partial_{k_{1,1}}\sigma_0(k_{1,1})\right),
\end{align*}
respectively. Thus, $\varphi_0=\ln\left(1-\bU_0^{(0,-1)}\right)$ provides a double-pole solution to the sinh--Gordon equation \eqref{sG}.
\end{example}

\begin{example}\label{E:Tz}
The third example is the Tzitzeica equation \eqref{Tz2}, as the first nontrivial example in the $A_{2r}^{(2)}$ class, namely the $r=1$ case.
In this case, there are two primitive cube roots of unity $\omega_3^{(1)}=\omega_3$ and $\omega_3^{(2)}=\omega_3^2$, where $\omega_3=\exp(2\pi\ri/3)$.
We consider measures
\begin{align*}
\mathrm{d}\lambda_1(k)=A_{1,1}\frac{1}{2\pi\mathrm{i}}\frac{1}{k-k_{1,1}}\rd k \quad \hbox{and} \quad
\mathrm{d}\lambda_2(k)=A_{2,1}\frac{1}{2\pi\mathrm{i}}\frac{1}{(k-k_{2,1})^2}\rd k,
\end{align*}
as particular cases of \eqref{Ar1:Pole}, which implies that there are two singularities on the $k$-plane, including
the first-order singularity $k_{1,1}$ located in the contour $\Gamma_1$ and the second-order singularity $k_{2,1}$ located in $\Gamma_2$.
Then equation \eqref{CN:Sol} implies that the Tzitzeica equation has its Cauchy matrix solution $\varphi_0=\ln\left(1-\bU_0^{(0,-1)}\right)$
with the expression of $\bU_0^{(0,-1)}$ given by
\begin{align*}
\bU_0^{(0,-1)}=(\tbr,\tbr')
\left[\bI+
\begin{pmatrix}
\bA& 0 \\
0& \bA'
\end{pmatrix}
\begin{pmatrix}
\bM^{(1,1)}& \bM^{(1,2)} \\
\bM^{(2,1)}& \bM^{(2,2)}
\end{pmatrix}
\right]^{-1}
\begin{pmatrix}
\bA& 0 \\
0& \bA'
\end{pmatrix}
\begin{pmatrix}
\bK'^{(-1)}& 0\\
0 & \bK^{(-1)}
\end{pmatrix}
\left(
\begin{array}{c}
\bs \\
\bs'
\end{array}
\right).
\end{align*}
Here $\bA$ and $\bA'$ are the $3\times 3$ diagonal matrices
\begin{align*}
\bA=\mathrm{Diag}\left(A_{1,1},A_{2,1},A_{2,1}\right) \quad \hbox{and} \quad
\bA'=\mathrm{Diag}\left(A_{1,1},-\oa_3^{(2)}A_{2,1},A_{2,1}\right).
\end{align*}
The $3$-component vectors $\br$, $\br'$, $\bs$ and $\bs'$ are defined by
\begin{align*}
\tbr=\left(\rho_0(k_{1,1}),\partial_{k_{2,1}}\rho_0(k_{2,1}),\rho_0(k_{2,1})\right),\quad
\tbr'=\left(\rho_0(-\oa_3^{(1)}k_{1,1}),\partial_{-\oa_3^{(2)}k_{2,1}}\rho_0(-\oa_3^{(2)}k_{2,1}),\rho_0(-\oa_3^{(2)}k_{2,1})\right),
\end{align*}
and
\begin{align*}
\tbs=\left(\sigma_0(-\oa_3^{(1)}k_{1,1}),\sigma_0(-\oa_3^{(2)}k_{2,1}),\partial_{k_{2,1}}\sigma_0(-\oa_3^{(2)}k_{2,1})\right),
\quad
\tbs'=\left(\sigma_0(k_{1,1}),\sigma_0(k_{2,1}),\partial_{k_{2,1}}\sigma_0(k_{2,1})\right),
\end{align*}
respectively. $\bK^{(-1)}$ and $\bK'^{(-1)}$ are $3\times3$ lower-triangle matrices
\begin{align*}
\bK^{(-1)}=
\begin{pmatrix}
k^{-1}_{1,1}& 0& 0\\
0&k^{-1}_{2,1}& 0 \\
0&-k^{-2}_{2,1} & k^{-1}_{2,1}
\end{pmatrix}
\quad \hbox{and} \quad
\bK'^{(-1)}=
\begin{pmatrix}
-\oa_3^{(2)}k_{1,1}^{-1}& 0& 0\\
0 & -\oa_3^{(1)}k_{2,1}^{-1}& 0 \\
0 & \oa_3^{(1)}k_{2,1}^{-2} & -\oa_3^{(1)}k_{2,1}^{-1}
\end{pmatrix},
\end{align*}
respectively.
In either $\bK^{(-1)}$ or $\bK'^{(-1)}$, we observe that the upper-left corner is a scalar, corresponding to the singularity $k_{1,1}$ of order $1$;
while the lower-right corner is a $2\times 2$ Jordan matrix, corresponding to the singularity $k_{2,1}$ of order $2$.
In other words, the two singularities results in the Cauchy matrix solution being the mixture of a simple-pole soliton and a double-pole soliton.
$\bM^{(\alpha,\beta)}$ for $\alpha,\beta\in\{1,2\}$ in the Cauchy matrix is of the form
\begin{align*}
\bM^{(\alpha,\beta)}=
\begin{pmatrix}
\bM^{(\alpha,\beta)}_{1,1;1,1} & \bM^{(\alpha,\beta)}_{1,1;2,1}\\
\bM^{(\alpha,\beta)}_{2,1;1,1} & \bM^{(\alpha,\beta)}_{2,1;2,1}
\end{pmatrix},
\end{align*}
in which $\bM^{(\alpha,\beta)}_{1,1;1,1}$ is a scalar,
$\bM^{(\alpha,\beta)}_{1,1;2,1}$ and $\bM^{(\alpha,\beta)}_{2,1;1,1}$ are $2$-component row and column vectors, respectively,
and $\bM^{(\alpha,\beta)}_{2,1;2,1}$ is a $2\times 2$ matrix.
When $\alpha=1$ and $\beta=1$, we have for $\bM^{(1,1)}$ the following blocks:
\begin{align*}
&\bM^{(1,1)}_{1,1;1,1}=\frac{\sigma_0(-\oa_3^{(1)}k_{1,1})\rho_0(k_{1,1})}{k_{1,1}-\oa_3^{(1)}k_{1,1}}, \\
&\bM^{(1,1)}_{1,1;2,1}=\sigma_0(-\oa_3^{(1)}k_{1,1})\left(\frac{1}{k_{2,1}-\oa_3^{(1)}k_{1,1}},\frac{-1}{(k_{2,1}-\oa_3^{(1)}k_{1,1})^2}\right)
\begin{pmatrix}
\partial_{k_{2,1}}\rho_0(k_{2,1})& \rho_0(k_{2,1}) \\
\rho_0(k_{2,1}) & 0
\end{pmatrix}, \\
&\bM^{(1,1)}_{2,1;1,1}=
\begin{pmatrix}
\sigma_0(-\oa_3^{(2)}k_{2,1})& 0 \\
\partial_{k_{2,1}}\sigma_0(-\oa_3^{(2)}k_{2,1}) & \sigma_0(-\oa_3^{(2)}k_{2,1})
\end{pmatrix}
\left(
\begin{array}{c}
\frac{1}{k_{1,1}-\oa_3^{(2)}k_{2,1}}\\
\frac{\oa_3^{(2)}}{(k_{1,1}-\oa_3^{(2)}k_{2,1})^2}
\end{array}
\right)\rho_0(k_{1,1}), \\
&\bM^{(1,1)}_{2,1;2,1}=
\begin{pmatrix}
\sigma_0(-\oa_3^{(2)}k_{2,1})& 0 \\
\partial_{k_{2,1}}\sigma_0(-\oa_3^{(2)}k_{2,1}) & \sigma_0(-\oa_3^{(2)}k_{2,1})
\end{pmatrix}
\begin{pmatrix}
\frac{1}{k_{2,1}-\oa_3^{(2)}k_{2,1}}& \frac{-1}{(k_{2,1}-\oa_3^{(2)}k_{2,1})^2}\\
\frac{\oa_3^{(2)}}{(k_{2,1}-\oa_3^{(2)}k_{2,1})^2} & \frac{-2\oa_3^{(2)}}{(k_{2,1}-\oa_3^{(2)}k_{2,1})^3}
\end{pmatrix}
\begin{pmatrix}
\partial_{k_{2,1}}\rho_0(k_{2,1})& \rho_0(k_{2,1}) \\
\rho_0(k_{2,1}) & 0
\end{pmatrix}.
\end{align*}
When $\alpha=1$ and $\beta=2$, the corresponding blocks in $\bM^{(1,2)}$ are as follows:
\begin{align*}
&\bM^{(1,2)}_{1,1;1,1}=\frac{\sigma_0(-\oa_3^{(1)}k_{1,1})\rho_0(-\oa_3^{(1)}k_{1,1})}{-\oa_3^{(1)}k_{1,1}-\oa_3^{(1)}k_{1,1}}, \\
&\bM^{(1,2)}_{1,1;2,1}=\sigma_0(-\oa_3^{(1)}k_{1,1})\left(\frac{1}{-\oa_3^{(1)}k_{2,1}-\oa_3^{(1)}k_{1,1}},\frac{-1}{(-\oa_3^{(1)}k_{2,1}-\oa_3^{(1)}k_{1,1})^2}\right)
\begin{pmatrix}
\partial_{-\oa_3^{(1)}k_{2,1}}\rho_0(-\oa_3^{(1)}k_{2,1})& \rho_0(-\oa_3^{(1)}k_{2,1}) \\
\rho_0(-\oa_3^{(1)}k_{2,1}) & 0
\end{pmatrix}, \\
&\bM^{(1,2)}_{2,1;1,1}=
\begin{pmatrix}
\sigma_0(-\oa_3^{(2)}k_{2,1})& 0 \\
\partial_{k_{2,1}}\sigma_0(-\oa_3^{(2)}k_{2,1}) & \sigma_0(-\oa_3^{(2)}k_{2,1})
\end{pmatrix}
\left(
\begin{array}{c}
\frac{1}{-\oa_3^{(2)}k_{1,1}-\oa_3^{(2)}k_{2,1}}\\
\frac{\oa_3^{(2)}}{(-\oa_3^{(2)}k_{1,1}-\oa_3^{(2)}k_{2,1})^2}
\end{array}
\right)\rho_0(-\oa_3^{(2)}k_{1,1}), \\
&\bM^{(1,2)}_{2,1;2,1}=
\begin{pmatrix}
\sigma_0(-\oa_3^{(2)}k_{2,1})& 0 \\
\partial_{k_{2,1}}\sigma_0(-\oa_3^{(2)}k_{2,1}) & \sigma_0(-\oa_3^{(2)}k_{2,1})
\end{pmatrix}
\begin{pmatrix}
\frac{1}{-\oa_3^{(2)}k_{2,1}-\oa_3^{(2)}k_{2,1}}& \frac{-1}{(-\oa_3^{(2)}k_{2,1}-\oa_3^{(2)}k_{2,1})^2}\\
\frac{\oa_3^{(2)}}{(-\oa_3^{(2)}k_{2,1}-\oa_3^{(2)}k_{2,1})^2} & \frac{-2\oa_3^{(2)}}{(-\oa_3^{(2)}k_{2,1}-\oa_3^{(2)}k_{2,1})^3}
\end{pmatrix}
\begin{pmatrix}
\partial_{-\oa_3^{(2)}k_{2,1}}\rho_0(-\oa_3^{(2)}k_{2,1})& \rho_0(-\oa_3^{(2)}k_{2,1}) \\
\rho_0(-\oa_3^{(2)}k_{2,1}) & 0
\end{pmatrix}.
\end{align*}
For $\alpha=2$ and $\beta=1$, $\bM^{(2,1)}$ takes its corresponding blocks as follows:
\begin{align*}
&\bM^{(2,1)}_{1,1;1,1}=\frac{\sigma_0(k_{1,1})\rho_0(k_{1,1})}{k_{1,1}+k_{1,1}}, \\
&\bM^{(2,1)}_{1,1;2,1}=\sigma_0(k_{1,1})\left(\frac{1}{k_{2,1}+k_{1,1}},\frac{-1}{(k_{2,1}+k_{1,1})^2}\right)
\begin{pmatrix}
\partial_{k_{2,1}}\rho_0(k_{2,1})& \rho_0(k_{2,1}) \\
\rho_0(k_{2,1}) & 0
\end{pmatrix}, \\
&\bM^{(2,1)}_{2,1;1,1}=
\begin{pmatrix}
\sigma_0(k_{2,1})& 0 \\
\partial_{k_{2,1}}\sigma_0(k_{2,1}) & \sigma_0(k_{2,1})
\end{pmatrix}
\left(
\begin{array}{c}
\frac{1}{k_{1,1}+k_{2,1}}\\
\frac{-1}{(k_{1,1}+k_{2,1})^2}
\end{array}
\right)\rho_0(k_{1,1}), \\
&\bM^{(2,1)}_{2,1;2,1}=
\begin{pmatrix}
\sigma_0(k_{2,1})& 0 \\
\partial_{k_{2,1}}\sigma_0(k_{2,1}) & \sigma_0(k_{2,1})
\end{pmatrix}
\begin{pmatrix}
\frac{1}{k_{2,1}+k_{2,1}}& \frac{-1}{(k_{2,1}+k_{2,1})^2}\\
\frac{-1}{(k_{2,1}+k_{2,1})^2} & \frac{2}{(k_{2,1}+k_{2,1})^3}
\end{pmatrix}
\begin{pmatrix}
\partial_{k_{2,1}}\rho_0(k_{2,1})& \rho_0(k_{2,1}) \\
\rho_0(k_{2,1}) & 0
\end{pmatrix}.
\end{align*}
For $\alpha=2$ and $\beta=1$, $\bM^{(2,2)}$ is composed of the following blocks:
\begin{align*}
&\bM^{(2,2)}_{1,1;1,1}=\frac{\sigma_0(k_{1,1})\rho_0(-\oa_3^{(1)}k_{1,1})}{-\oa_3^{(1)}k_{1,1}+k_{1,1}}, \\
&\bM^{(2,2)}_{1,1;2,1}=\sigma_0(k_{1,1})\left(\frac{1}{-\oa_3^{(1)}k_{2,1}+k_{1,1}},\frac{-1}{(-\oa_3^{(1)}k_{2,1}+k_{1,1})^2}\right)
\begin{pmatrix}
\partial_{-\oa_3^{(1)}k_{2,1}}\rho_0(-\oa_3^{(1)}k_{2,1})& \rho_0(-\oa_3^{(1)}k_{2,1}) \\
\rho_0(-\oa_3^{(1)}k_{2,1}) & 0
\end{pmatrix}, \\
&\bM^{(2,2)}_{2,1;1,1}=
\begin{pmatrix}
\sigma_0(k_{2,1})& 0 \\
\partial_{k_{2,1}}\sigma_0(k_{2,1}) & \sigma_0(k_{2,1})
\end{pmatrix}
\left(
\begin{array}{c}
\frac{1}{-\oa_3^{(2)}k_{1,1}+k_{2,1}}\\
\frac{-1}{(-\oa_3^{(2)}k_{1,1}+k_{2,1})^2}
\end{array}
\right)\rho_0(-\oa_3^{(2)}k_{1,1}), \\
&\bM^{(2,2)}_{2,1;2,1}=
\begin{pmatrix}
\sigma_0(k_{2,1})& 0 \\
\partial_{k_{2,1}}\sigma_0(k_{2,1}) & \sigma_0(k_{2,1})
\end{pmatrix}
\begin{pmatrix}
\frac{1}{-\oa_3^{(2)}k_{2,1}+k_{2,1}}& \frac{-1}{(-\oa_3^{(2)}k_{2,1}+k_{2,1})^2}\\
\frac{-1}{(-\oa_3^{(2)}k_{2,1}+k_{2,1})^2} & \frac{2}{(-\oa_3^{(2)}k_{2,1}+k_{2,1})^3}
\end{pmatrix}
\begin{pmatrix}
\partial_{-\oa_3^{(2)}k_{2,1}}\rho_0(-\oa_3^{(2)}k_{2,1})& \rho_0(-\oa_3^{(2)}k_{2,1}) \\
\rho_0(-\oa_3^{(2)}k_{2,1}) & 0
\end{pmatrix}.
\end{align*}
We comment that the blocks $\bM^{(\alpha,\beta)}_{1,1;1,1}$ and $\bM^{(\alpha,\beta)}_{2,1;2,1}$
describe the behaviour of the simple-pole soliton and the double-pole soliton themselves;
while the blocks $\bM^{(\alpha,\beta)}_{1,1;2,1}$ and $\bM^{(\alpha,\beta)}_{2,1;1,1}$ explain the interaction between the two solitons.
\end{example}

\section{Concluding remarks}\label{S:Concl}

We established the \ac{DL} schemes for the \ac{2D} Toda-type equations associated with the infinite-dimensional algebras $A_\infty$, $B_\infty$ and $C_\infty$,
as well as the Kac--Moody algebras $A_{r}^{(1)}$, $A_{2r}^{(2)}$, $C_r^{(1)}$ and $D_{r+1}^{(2)}$.
This was realised by associating the \ac{2DTL}s with the linear integral equations taking the form of \eqref{Linear}.
The starting point was the linear integral equation \eqref{Lineara} and its adjoint \eqref{Linearb} with arbitrary integration measure and domain,
whose nonlinearisation resulted in the $A_\infty$-type \ac{2DTL};
while the \ac{2D} Toda equations of other types were obtained by imposing various constraints on the measure and the domain in the $A_\infty$ class.
The linear integral equations also helped to rediscover the Lax scheme (including both Lax pair and its adjoint) and the bilinear formulation in the \ac{2D} Toda theory,
given by Fordy and Gibbons \cite{FG80,FG83}, Nimmo and Willox \cite{NW97}, as well as the Kyoto school \cite{JM83,UT84}.

The key point in the construction of the \ac{2D} Toda-type equations is the introduction of the infinite matrix $\bU$.
This allows us to transfer the discrete and continuous dynamical relations to purely algebraic ones.
Then the problem of constructing integrable systems turns to be searching for identities of the entries of the infinite matrix.
In other words, the \ac{DL} approach is a method to construct nonlinear equations directly.
There are also other methods to construct the \ac{2D} Toda systems such as the Lax scheme and the bilinear approach,
but the motivation seems different compared with the \ac{DL}, to the best of the authors' knowledge.
For example, the starting point in the fermionic approach (i.e. the bilinear method) is a bilinear identity,
which can generate bilinear \ac{2D} Toda equations as the main object, see \cite{JM83,UT84}.
As for the Lax scheme for the \ac{2DTL}s (see e.g. \cite{FG83}),
the key step is to construct linear spectral problems corresponding to various Kac--Moody algebras,
while the nonlinear equations arise as the compatibility conditions of the corresponing linear problems.

The linear integral equations theoretically provide a very large class of solutions for nonlinear integral equations.
This is because the formal solution obtained from the scheme, namely \eqref{Potential},
is an integral defined on a very general integration domain in the complex space in terms of the spectral parameter(s), compared with other methods.
Our main achievement in this paper is that we successfully constructed direct linearising type solution for each class of the \ac{2D} Toda-type equation.

As a degenerate case of the direct linearising type solution, we presented the general formula of the Cauchy matrix type solution for each class of \ac{2DTL},
by imposing the condition that the integration is defined on the domain containing an arbitrary number of singularities of arbitrary order.
The solutions of $A_\infty$- and $A_1^{(1)}$-types coincide with those obtained from the operator approach and the so-called generalised Cauchy matrix approach,
see e.g. \cite{Sch05,Sch10,NS20} and also \cite{NAH09,ZZ13,XZZ14}.
Here we extended these results to nonlinear integrable systems associated with other Lie algebras.

The procedure of deriving Cauchy matrix type solutions can actually reduce the formal tau functions in the form of \eqref{tau} to the corresponding degenerate cases,
resulting in determinant type solutions for the bilinear forms of the \ac{2D} Toda systems.
These solutions are closely related to the so-called Grammian solutions obtained from the well-known Hirota method, see e.g. \cite{Hir04}.
We also note that \ac{DL} in some cases induces more general solutions, compared with Grammian solutions.
For instance, the numbers of the singularities on the $k$- and $k'$-plane are not necessarily equal in our result for the $A_\infty$-type \ac{2DTL}.
This simultaneously results in the existence of the finite matrix $\bA$ in \eqref{A:Sol}, see also example \ref{E:2DTL}.
In fact, the coefficient matrix $\bA$ plays a crucial role in the classification of solitons, explaining a variety of soliton behaviours, see e.g. \cite{KW11,KW13,KW14,BMW17}.

The \ac{DL} scheme of the \ac{2DTL} of $D_\infty$-type and its reduced systems, as well as the equations associated with the exceptional Lie algebras,
are not covered by the scheme in this paper.
Searching for other classes of exact solutions for the \ac{2D} Toda-type equations from the linear integral equations is also an interesting problem,
but it then requires nontrivial spectral analysis for the linear integral equations. This remains for our future work.

\section*{Acknowledgments}
This project was supported by the National Natural Science Foundation of China (grant no. 11901198) and Shanghai Pujiang Program (grant no. 19PJ1403200).
WF was also partially sponsored by the Science and Technology Commission of Shanghai Municipality (grant no. 18dz2271000)
as well as the Fundamental Research Funds for the Central Universities.

\renewcommand{\bibname}{References}
\bibliography{References}
\bibliographystyle{plain}

\end{document}